\documentclass[12pt]{article}

\usepackage{graphicx}
\usepackage{amsmath,amssymb}
\usepackage{bm}
\usepackage{graphicx, color}
\usepackage{wrapfig}
\begin{document}
\title{
\begin{flushright}
\ \\*[-80pt] 
\begin{minipage}{0.2\linewidth}
\normalsize
\end{minipage}
\end{flushright}
{\Large \bf 
Testing the minimal $S_4$ model of neutrinos \\
with the Dirac and Majorana phases 
\\*[20pt]}}

\author{ 
\centerline{
Yusuke~Shimizu$^{1,}$\footnote{E-mail address: shimizu@muse.sc.niigata-u.ac.jp},
~Morimitsu~Tanimoto$^{2,}$\footnote{E-mail address: tanimoto@muse.sc.niigata-u.ac.jp} }
\\*[20pt]
\centerline{
\begin{minipage}{\linewidth}
\begin{center}
$^1${\it \normalsize
Graduate~School~of~Science~and~Technology,~Niigata~University, \\ 
Niigata~950-2181,~Japan } \\
$^2${\it \normalsize
Department of Physics, Niigata University,~Niigata 950-2181, Japan } 
\end{center}
\end{minipage}}
\\*[50pt]}

\date{
\centerline{\small \bf Abstract}
\begin{minipage}{0.9\linewidth}
\medskip 
\medskip 
\small
We propose two new simple lepton flavor models 
in the framework of the $S_4$ flavor symmetry. 
The neutrino mass matrices, which are given by two complex parameters, lead to the inverted mass hierarchy. 
The charged lepton mass matrix has the 1-2 lepton flavor mixing, 
which gives the non-vanishing reactor angle $\theta_{13}$. 
These models predict the Dirac phase and the Majorana phases, which are testable 
in the future experiments. 
The predicted magnitudes of the effective neutrino mass for the neutrino-less double beta decay 
are in the regions as $32~\text{meV}\lesssim |m_{ee}|\lesssim 49~\text{meV}$ 
and $34~\text{meV}\lesssim |m_{ee}|\lesssim 59~\text{meV}$, respectively. 
These values are close to the expected reaches of the coming experiments. 
The total sum of the neutrino masses are predicted in both models as $0.0952~\text{eV}\lesssim \sum m_i\lesssim 0.101~\text{eV}$ 
and $0.150~\text{eV}\lesssim \sum m_i\lesssim 0.160~\text{eV}$, respectively.
\end{minipage}
}

\begin{titlepage}
\maketitle
\thispagestyle{empty}
\end{titlepage}

\section{Introduction}
The neutrino experiments are going on the new step to reveal the CP violations
in the lepton sector. If the neutrinos are the Majorana particles,
 there are one Dirac phase and two Majorana phases~\cite{Bilenky:1980cx,Schechter:1980gr,Doi:1980yb},
 which are sources of the CP violation.
The T2K experiment, which has already confirmed the neutrino oscillation in the $\nu_\mu \to\nu_e$ appearance events \cite{Abe:2013hdq},
 provides us the new information of the CP violating Dirac phase by combining  the data of reactor experiments~\cite{An:2012eh}-\cite{Abe:2014lus}.
On the other hand, the neutrino-less double beta decay ($0\nu \beta \beta$) experiments 
lower gradually the upper-bound of the effective neutrino mass $m_{ee}$
around ${\cal O}(100)$~meV.  If the neutrino mass hierarchy is the inverted one, one can expect
to observe the $0\nu \beta \beta $  in the near future. Its magnitude depends on  the CP violating Majorana phases.
Thus, the three CP violating phases can be observed in progressive neutrino experiments.
Those  CP violating phases provide us the important information to test the flavor model
of neutrinos. There appear many works to test models
with the CP violating Dirac and the Majorana phases of neutrinos in addition to the
mixing angle sum rules~\cite{Kang:2000sb}-\cite{Kang:2015xfa}.

The recent experimental data of neutrinos give us a big hint of the flavor symmetry.
Actually, the lepton flavor structure  has been discussed
in the framework of the flavor symmetry.
Before the reactor experiments reported 
the non-zero value of $\theta _{13}$,  there appears  a paradigm of
 "tri-bimaximal mixing" (TBM)~\cite{Harrison:2002er,Harrison:2002kp}, 
which is a simple mixing pattern for leptons and can be easily derived from flavor symmetries. 
Some authors  succeeded to obtain  the TBM in the $A_4$ models~\cite{Ma:2001dn}-\cite{deMedeirosVarzielas:2005qg}.
After those successes, the non-Abelian discrete groups are center of attention at the flavor symmetry~\cite{Altarelli:2010gt}-\cite{King:2014nza}.
The other groups were also examined to give the TBM~\cite{Mohapatra:2006pu}-\cite{Altarelli:2009gn}.
  The deviations from the TBM  were estimated in many works~\cite{Xing:2002sw},~\cite{Xing:2006ms}-\cite{Bazzocchi:2012st}. 
The observation of the non-vanishing $\theta_{13}$ enables the detail studies of flavor models 
in the context of the sum rules of mixing angles~\cite{King:2014nza},~\cite{Acosta:2012qf}-\cite{King:2013xba}.
The neutrino CP violation is also discussed in the context of the generalized CP symmetry~\cite{Branco:2011zb,Grimus:2003yn,Chen:2009gf}.
The  non-Abelian discrete groups with the CP symmetry have  predicted the magnitude of the
CP violating phases~\cite{Feruglio:2013hia}-\cite{Li:2015jxa}.
Furthermore, the neutrino mixing angles have been predicted linking with the 
quark mixing matrix 
 by using some GUT models~\cite{Bazzocchi:2008sp}-\cite{Yao:2015dwa}.
Thus, the flavor models in the lepton sector confront the neutrino experimental data of CP violating phases.
 
However, the flavor models do not always give the unique predictions 
since they have many parameters. Therefore, it is desirable to build a model 
with the small number of parameters for testing it.
This attempt was proposed with Occam's Razor~\cite{Harigaya:2012bw}, where the inverted mass hierarchy is predicted.
In this paper, we build two  neutrino flavor models with the small number of parameters
as much as possible
in the framework of the indirect approach with the $S_4$ flavor symmetry.
The neutrino mass matrix
is given by two complex parameters. The charged lepton mass matrix  gives the non-vanishing reactor angle $\theta_{13}$. 
The models lead to the inverted mass hierarchy, and predict the Dirac phase and the Majorana phases. 
Our predicted effective neutrino mass of the $0\nu \beta \beta $  is  correlated with the Dirac phase.

In section 2, we propose two simple $S_4$ models, and discuss their implications.
In section~3, we present the numerical analyses of the CP violating phases
as well as the mixing angles, and then  predict the effective neutrino mass for 
the $0\nu \beta \beta $. 
The Section 4 is devoted to discussions and summary. 
In Appendices, we present the multiplication rules of the $S_4$ group, and
 the neutrino mass matrix in the relevant basis of  neutrinos.


\section{$S_4$ flavor models}
Let us build  simple lepton flavor models with the $S_4$ group by the indirect approach
of the flavor symmetry. We  obtain the lepton mass matrices
by introducing the relevant flavons and assuming the alignment of their vacuum expectation values (VEVs). 
The advantage of the $S_4$ group includes both the doublet and the triplets as the fundamental representations. 
Therefore, the $S_4$ flavor symmetry was easily extended to the quark sector \cite{Ishimori:2008fi,Ishimori:2010xk}.
We obtain two models with  the inverted neutrino mass hierarchy, which give two cases
 of  the lightest neutrino mass $m_3=0$ and $m_3\not =0$, respectively. 
The particle assignments are same in both cases, while the vacuum alignments of flavon for the neutrino sector are different in each model.

\subsection{$S_4$ flavor model for $m_3=0$}
We present a model with the $S_4$ flavor symmetry in the case of the vanishing lightest neutrino mass. 
The particle assignments are shown in Table~\ref{tab:assignment}. 
These assignments are similar to the model in Refs.~\cite{Ishimori:2008fi,Ishimori:2010xk} except for flavon fields. 
Under the $S_4$ group, 
the left-handed lepton doublet $\bar L=(\bar L_e, \bar L_\mu, \bar L_\tau)$ are assumed to transform
as the triplet, while the right-handed  charged lepton are assigned
to the doublet $\ell_R=(e_R, \mu_R)$ and the singlet $\tau_R$.
The right-handed neutrinos are also assigned to the doublet $N^c_R=(N_{eR}^c, N_{\mu R}^c)$ and the singlet $N^c_{\tau R}$.
The $Z_4$ charge is assigned relevantly to the leptons. 
In order to get the natural hierarchy between the muon and tauon masses, the Froggatt-Nielsen mechanism~\cite{Froggatt:1978nt} is introduced
as an additional $U(1)_{FN}$ flavor symmetry, where $\Theta $ denotes the Froggatt-Nielsen flavon. 
On the other hand,  the Higgs doublet $H$ is assigned to the $S_4$ singlet.
The gauge singlet flavons are assigned to the triplet or triplet-prime, which have
different $Z_4$ charges as seen in Table 1.

We can now write down the $S_4\times Z_4\times U(1)_{FN}$ invariant Lagrangian for the  Yukawa interaction in terms of the $S_4$ cutoff scale $\Lambda$ and
the $U(1)_{FN}$ cutoff scale $\bar\Lambda$  as follows:
\begin{align}
\mathcal{L}_Y&=y_\ell \bar Ll_RH\phi _\ell \Theta ^2/(\Lambda \bar \Lambda ^2)
+y_\ell '\bar Ll_RH\phi _\ell '\Theta ^2/(\Lambda \bar \Lambda ^2)+y_\ell ''\bar L\tau _RH\phi _\ell '' /\Lambda \nonumber \\
&+y_D\bar LN_R^c\widetilde H\phi _\nu /\Lambda +y_D'\bar LN_{\tau R}^c\widetilde H\phi _\nu '/\Lambda \nonumber \\
&+M_1N_R^cN_R^c+M_2N_{\tau R}^cN_{\tau R}^c \ ,
\label{lagrangian}
\end{align}
where $y$'s are Yukawa couplings, $M_1$ and $M_2$ are the Majorana masses, 
 and $\widetilde H=i\tau _2H^*$.
In this setup, we discuss the lepton mass matrices.
\begin{table}[hbtp]
\begin{center}
\begin{tabular}{|c|ccccc|c|cccccc|}
\hline 
& $\bar L$ & $l_R$ & $\tau _R$ & $N_R^c$ & $N_{\tau R}^c$ & $H$ 
& $\phi _\nu $ & $\phi _\nu '$ & $\phi _\ell $ & $\phi _\ell '$ & $\phi _\ell ''$ & $\Theta $ \\
\hline 
$SU(2)$ & ${\bf 2}$ & ${\bf 1}$ & ${\bf 1}$ & ${\bf 1}$ & ${\bf 1}$ & ${\bf 2}$ & 
${\bf 1}$ & ${\bf 1}$ & ${\bf 1}$ & ${\bf 1}$ & ${\bf 1}$ & ${\bf 1}$ \\
$S_4$ & ${\bf 3}$ & ${\bf 2}$ & ${\bf 1}$ & ${\bf 2}$ & ${\bf 1}$ & ${\bf 1}$ 
& ${\bf 3}$ & ${\bf 3}$ & ${\bf 3}$ & ${\bf 3}'$ & ${\bf 3}$ & ${\bf 1}$ \\
$Z_4$ & $1$ & $i$ & $-i$ & $-1$ & $1$ & $1$ 
& $-1$ & $1$ & $-i$ & $-i$ & $i$ & $1$ \\
$U(1)_{FN}$ & $0$ & $\ell +2$ & $0$ & $0$ & $0$ & $0$ 
& $0$ & $0$ & $-\ell $ & $-\ell $ & $0$ & $-1$ \\
\hline 
\end{tabular}
\caption{The assignments of leptons, Higgs and flavons.}
\label{tab:assignment}
\end{center}
\end{table}

Let us begin with discussing the neutrino sector.
In order to desirable mass matrices, the relevant VEV alignments are required.
We will show the potential analysis to derive  the VEV alignments in subsection 2.3.

We take the  VEVs of the relevant flavons and VEV alignments as:
\begin{equation}
\langle H\rangle =v,\quad \langle \phi _\nu \rangle = v_\nu (1,0,0),\quad 
\langle \phi _\nu '\rangle = v_\nu '(1,1,1).
\label{alignments}
\end{equation}
By using the multiplication rules in Appendix A,
we obtain  the Dirac neutrino mass matrix $M_D$  as
\begin{equation}
M_D=
\begin{pmatrix}
0 & -\frac{2}{\sqrt{6}}y_Dv_\nu & y_D'v_\nu ' \\
0 & 0 & y_D'v_\nu ' \\
0 & 0 & y_D'v_\nu '
\end{pmatrix}\frac{v}{\Lambda } \ ,
\end{equation}
and the right-handed Majorana neutrino mass matrix $M_N$ as
\begin{equation}
M_N=
\begin{pmatrix}
M_1 & 0 & 0 \\
0 & M_1 & 0 \\
0 & 0 & M_2
\end{pmatrix} \ .
\label{Majoranamatrix}
\end{equation}
By using the seesaw mechanism, the left-handed Majorana neutrino mass matrix $M_\nu $ is given as
\begin{equation}
M_\nu =
\begin{pmatrix}
a+b & b & b \\
b & b & b \\
b & b & b
\end{pmatrix}, \qquad a=\frac{2(y_Dv_\nu v)^2}{3M_1\Lambda ^2},\qquad b=\frac{(y_D'v_\nu 'v)^2}{M_2\Lambda ^2} \ ,
\label{model1}
\end{equation}
where $a$ and $b$ are complex parameters. 
In order to compare with the TBM, we move  to the TBM base.
Then,  we can easily seen the family structure of the 
neutrino mass matrix. After rotating $M_\nu$ with the TBM matrix $V_\text{TBM}$:  
\begin{equation}
V_\text{TBM}=
\begin{pmatrix}
\frac{2}{\sqrt{6}} & \frac{1}{\sqrt{3}} & 0 \\
-\frac{1}{\sqrt{6}} & \frac{1}{\sqrt{3}} & -\frac{1}{\sqrt{2}} \\
-\frac{1}{\sqrt{6}} & \frac{1}{\sqrt{3}} & \frac{1}{\sqrt{2}}
\end{pmatrix} \ ,
\end{equation}
the  neutrino mass matrix turns to
\begin{equation}
M_\nu =V_\text{TBM}
\begin{pmatrix}
\frac{2}{3}a & \frac{\sqrt{2}}{3}a & 0 \\
\frac{\sqrt{2}}{3}a & \frac{1}{3}a+3b & 0 \\
0 & 0 & 0
\end{pmatrix}V_\text{TBM}^T \ .
\end{equation}
This mass matrix gives us the inverted mass hierarchy with the one vanishing  mass.
The flavor mixing is deviated from the TBM only in the rotation of the (1-2) axis.

In order to get the mass eigenvalues and the flavor mixing angles of neutrinos, 
we examine $M_\nu ^\dagger M_\nu $ as 
\begin{align}
M_\nu ^\dagger M_\nu &=V_\text{TBM}
\begin{pmatrix}
\frac{2}{3}|a|^2 & \frac{\sqrt{2}}{3}|a|^2+\sqrt{2}a^*b & 0 \\
\frac{\sqrt{2}}{3}|a|^2+\sqrt{2}ab^* & \frac{1}{3}|a|^2+9|b|^2+ab^*+a^*b & 0 \\
0 & 0 & 0
\end{pmatrix}V_\text{TBM}^T \nonumber \\
&=V_\text{TBM}
\begin{pmatrix}
\frac{2}{3}|a|^2 & \frac{\sqrt{2}}{3}|a|^2+\sqrt{2}|a||b|e^{-i\varphi } & 0 \\
\frac{\sqrt{2}}{3}|a|^2+\sqrt{2}|a||b|e^{i\varphi } & \frac{1}{3}|a|^2+9|b|^2+2|a||b|\cos \varphi  & 0 \\
0 & 0 & 0
\end{pmatrix}V_\text{TBM}^T \ ,
\end{align}
where $a=|a|e^{i\varphi _a}$, $b=|b|e^{i\varphi _b}$, and $\varphi=\varphi_a-\varphi_b $~. 
Hereafter, we define $|a|\equiv a$ and $|b|\equiv b$ for simplicity. 
Then, the neutrino mass eigenvalues are 
\begin{align}
m_1^2&=\frac{1}{2}\left [a^2+9b^2+2ab\cos \varphi -\sqrt{\left (a^2+9 b^2+2ab\cos \varphi \right )^2-16a^2b^2}\right ], \nonumber \\
m_2^2&=\frac{1}{2}\left [a^2+9b^2+2ab\cos \varphi +\sqrt{\left (a^2+9 b^2+2ab\cos \varphi \right )^2-16a^2b^2}\right ], \nonumber \\
m_3^2&=0 \ .
\end{align}
Therefore, the neutrino mass squared differences are given as 
\begin{align}
\Delta m_{\text{atm}}^2&=m_2^2-m_3^2=\frac{1}{2}\left [a^2+9b^2+2ab\cos \varphi +\sqrt{\left (a^2+9 b^2+2ab\cos \varphi \right )^2-16a^2b^2}\right ], \nonumber \\
\Delta m_{\text{sol}}^2&=m_2^2-m_1^2=\sqrt{\left (a^2+9 b^2+2ab\cos \varphi \right )^2-16a^2b^2}~,
\label{mass-square-differences}
\end{align}
and neutrino mixing matrix $U_\nu $ is 
\begin{equation}
U_\nu =V_\text{TBM}
\begin{pmatrix}
\cos \eta & e^{-i\psi }\sin \eta & 0 \\
-e^{i\psi }\sin \eta & \cos \eta & 0 \\
0 & 0 & 1
\end{pmatrix} \ ,
\label{neutrino-mixing-matrix}
\end{equation}
where
\begin{equation}
\tan 2\eta =\frac{2\sqrt{2}a\sqrt{a^2+9b^2+6ab\cos \varphi }}{27b^2-a^2+6ab\cos \varphi }~, 
\qquad \tan \psi =-\frac{3b\sin \varphi }{a+3b\cos \varphi }~.
\label{tan2eta-tanpsi}
\end{equation}
Thus, the neutrino mass hierarchy is only inverted one 
and the neutrino mixing angles are determined by two neutrino mass squared differences and relative phase $\varphi$ in our model.

Next, we consider the charged lepton sector.
Taking VEV and VEV alignments as 
\footnote{We take the VEV of $\phi _{\ell }$ as $(v_{\ell 1},v_{\ell 2},0)$ for simplicity. 
As the alternative choice, we can introduce one more field $\tilde \phi _{\ell }$, which charge assignment of $S_4$ and $Z_4$ is same as $\phi _{\ell }$, 
and take VEV alignments as $\langle \phi _{\ell }\rangle =(v_{\ell 1},0,0)$, $\langle \tilde \phi _{\ell }\rangle =(0,v_{\ell 2},0).$
By using these VEV alignments, we obtain the same mass matrix in Eq.~(\ref{chargedlepton-massmatrix-other}).} 
\begin{equation}
\langle \Theta \rangle =\theta ,\quad \langle \phi _\ell \rangle =(v_{\ell 1},v_{\ell 2},0),\quad \langle \phi _\ell '\rangle =v_\ell '(1,0,0),\quad 
\langle \phi _\ell ''\rangle =v_\ell ''(0,0,1),
\end{equation}
the charged lepton mass matrix $M_\ell $ is given as
\begin{equation}
M_\ell =
\begin{pmatrix}
\frac{2}{\sqrt{6}}c_\ell & -\frac{2}{\sqrt{6}}a_\ell & 0 \\
\frac{1}{\sqrt{2}}b_\ell & \frac{1}{\sqrt{6}}b_\ell & 0 \\
0 & 0 & d_\ell 
\end{pmatrix},
\label{chargedlepton-massmatrix-other}
\end{equation}
where 
\begin{equation}
a_\ell =y_\ell v_{\ell 1}v \frac{\theta ^2}{\Lambda \bar\Lambda^{2}},
\quad 
b_\ell =y_\ell v_{\ell 2}v \frac{\theta ^2}{\Lambda \bar\Lambda^{2}},
\quad 
c_\ell =y_\ell 'v_\ell 'v \frac{\theta ^2}{\Lambda \bar\Lambda^{2}},
\quad 
d_\ell =y_\ell ''v_\ell ''v \frac{1}{\Lambda} .
\end{equation}
In order to reproduce the relevant mass hierarchy between 
the muon and the tauon, we take $\theta/\bar\Lambda$ as the Cabibbo angle $0.22$ approximately.

The left-handed mixing of charged leptons is given by  investigating  $M_\ell M_\ell ^\dagger $:
\begin{align}
M_\ell M_\ell ^\dagger &=
\begin{pmatrix}
\frac{2}{3}(|a_\ell |^2+|c_\ell |^2) & -\frac{1}{3}(a_\ell -\sqrt{3}c_\ell )b_\ell ^*& 0 \\
-\frac{1}{3}(a_\ell ^*-\sqrt{3}c_\ell ^*)b_\ell & \frac{2}{3}|b_\ell |^2 & 0 \\
0 & 0 & |d_\ell |^2
\end{pmatrix} \nonumber \\
&=
\begin{pmatrix}
\frac{2}{3}(|a_\ell |^2+|c_\ell |^2) & -\frac{1}{3}(|a_\ell ||b_\ell |e^{i\varphi _{ab}}-\sqrt{3}|b_\ell ||c_\ell |e^{-i\varphi _{bc}} )& 0 \\
-\frac{1}{3}(|a_\ell ||b_\ell |e^{-i\varphi _{ab}}-\sqrt{3}|b_\ell ||c_\ell |e^{i\varphi _{bc}})& \frac{2}{3}|b_\ell |^2 & 0 \\
0 & 0 & |d_\ell |^2
\end{pmatrix}~,
\end{align}
where $a_\ell =|a_\ell |e^{i\varphi _{a_\ell }}$, $b_\ell =|b_\ell |e^{i\varphi _{b_\ell }}$, $c_\ell =|c_\ell |e^{i\varphi _{c_\ell }}$, and 
$\varphi _{ab}=\varphi_{a_\ell}-\varphi_{b_\ell}$,
 $\varphi_{bc}=\varphi _{b_\ell }-\varphi _{c_\ell }$. 
Hereafter, we define $|a_\ell |\equiv a_\ell $, $|b_\ell |\equiv b_\ell $, and $|c_\ell |\equiv c_\ell $ for simplicity.
Then, the charged lepton masses 
are given as 
\begin{align}
m_e^2&=\frac{1}{3}\left [a_\ell ^2+b_\ell ^2+c_\ell ^2-\sqrt{\left (a_\ell ^2+b_\ell ^2+c_\ell ^2\right )^2
-b_\ell ^2\left (3a_\ell ^2+c_\ell ^2+2\sqrt{3}a_\ell c_\ell \cos \varphi _\ell \right )}\right ], \nonumber \\
m_\mu ^2&=\frac{1}{3}\left [a_\ell ^2+b_\ell ^2+c_\ell ^2+\sqrt{\left (a_\ell ^2+b_\ell ^2+c_\ell ^2\right )^2
-b_\ell ^2\left (3a_\ell ^2+c_\ell ^2+2\sqrt{3}a_\ell c_\ell \cos \varphi _\ell \right )}\right ], \nonumber \\
m_\tau ^2&=d_\ell ^2~,
\label{eq:chargedleptonmasses}
\end{align}
where $\varphi _\ell = \varphi _{ab}+\varphi _{bc}$~.

  The $U(1)_{FN}$ symmetry guarantees the mass hierarchy of the muon and the tauon.
 Now, we discuss  the electron and muon masses. 
If we assume  $b_\ell \gg a_\ell, c_\ell$  in Eq.~(\ref{chargedlepton-massmatrix-other}), the electron and muon masses in Eq.~(\ref{eq:chargedleptonmasses}) are approximately written as
\begin{align}
m_e^2&\simeq \frac{1}{3}\left (\frac{3}{2}a_\ell ^2+\frac{1}{2}c_\ell ^2+\sqrt{3}a_\ell c_\ell \cos \varphi _\ell \right ), \nonumber \\
m_\mu ^2&\simeq \frac{1}{3}\left (2b_\ell ^2+\frac{1}{2}a_\ell ^2+\frac{3}{2}c_\ell ^2-\sqrt{3}a_\ell c_\ell \cos \varphi _\ell \right ).
\end{align}
Then,  the muon mass  is $\sqrt{2/3} b_\ell$ and 
 the  electron  mass is reproduced by tuning   $a_\ell$ and $c_\ell$.

On the other hand, the left-handed charged lepton mixing matrix $U_\ell $ is 
\begin{equation}
U_\ell =
\begin{pmatrix}
\cos \lambda & e^{-i\psi _\ell }\sin \lambda & 0 \\
-e^{i\psi _\ell }\sin \lambda & \cos \lambda & 0 \\
0 & 0 & 1
\end{pmatrix},
\end{equation}
where
\begin{equation}
\tan 2\lambda =\frac{b_\ell \sqrt{a_\ell ^2+3c_\ell ^2-2\sqrt{3}a_\ell c_\ell \cos \varphi _\ell }}{a_\ell ^2-b_\ell ^2+c_\ell ^2}~, \quad 
\tan \psi _\ell =\frac{a_\ell \sin \varphi _{ab}+\sqrt{3}c_\ell \sin \varphi _{bc}}{a_\ell \cos \varphi _{ab}-\sqrt{3}c_\ell \cos \varphi _{bc}}~.
\end{equation}
In the charged lepton sector, there are  four real parameters ($a_\ell $, $b_\ell $, $c_\ell $, $d_\ell $) and two phases. 
After inputting charged lepton masses, there remains two  independent parameters,
which are the  mixing angle $\lambda$  and the  CP violating phase $\psi_\ell$. 
Those free parameters are adjusted to the experimental data of the lepton mixing matrix, 
namely Pontecorvo-Maki-Nakagawa-Sakata (PMNS) matrix $U_\text{PMNS}$~\cite{Maki:1962mu,Pontecorvo:1967fh} as  
\begin{equation}
U_\text{PMNS}=U_\ell ^\dagger U_\nu =
\begin{pmatrix}
\cos \lambda & -e^{-i\psi _\ell }\sin \lambda & 0 \\
e^{i\psi _\ell }\sin \lambda & \cos \lambda & 0 \\
0 & 0 & 1
\end{pmatrix}
V_\text{TBM}
\begin{pmatrix}
\cos \eta & e^{-i\psi }\sin \eta & 0 \\
-e^{i\psi }\sin \eta & \cos \eta & 0 \\
0 & 0 & 1
\end{pmatrix}.
\label{PMNS-matrix}
\end{equation}
The mixing matrix elements  are written  as
\begin{align}
U_{e1}&=\frac{\cos \eta \left (2\cos \lambda +e^{-i\psi _\ell }\sin \lambda \right )}{\sqrt{6}}
+\frac{e^{i\psi }\sin \eta \left (-\cos \lambda +e^{-i\psi _\ell }\sin \lambda \right )}{\sqrt{3}}~, \nonumber \\
U_{e2}&=\frac{\cos \eta \left (\cos \lambda -e^{-i\psi _\ell }\sin \lambda \right )}{\sqrt{3}}
+\frac{e^{-i(\psi +\psi _\ell )}\sin \eta \left (2e^{i\psi _\ell }\cos \lambda +\sin \lambda \right )}{\sqrt{6}}~, \nonumber \\
U_{\mu 1}&=\frac{\cos \eta \left (-\cos \lambda +2 e^{i\psi _\ell }\sin \lambda \right )}{\sqrt{6}}
-\frac{e^{i\psi }\sin \eta \left (\cos \lambda +e^{i\psi _\ell }\sin \lambda \right )}{\sqrt{3}}~, \nonumber \\
U_{\mu 2}&=\frac{\cos \eta \left (\cos \lambda +e^{i\psi _\ell }\sin \lambda \right )}{\sqrt{3}}
+\frac{e^{-i\psi }\sin \eta \left (-\cos \lambda +2 e^{i\psi _\ell }\sin \lambda \right )}{\sqrt{6}}~, \nonumber \\
U_{\tau 1}&=-\frac{\cos \eta }{\sqrt{6}}-\frac{e^{i \psi }\sin \eta }{\sqrt{3}}~, \qquad 
U_{\tau 2}=\frac{\cos \eta }{\sqrt{3}}-\frac{e^{-i\psi }\sin \eta }{\sqrt{6}}~, \nonumber \\
U_{e3}&=\frac{e^{-i\psi _\ell }\sin \lambda }{\sqrt{2}}~,\qquad 
U_{\mu 3}=-\frac{\cos \lambda }{\sqrt{2}}~,\qquad U_{\tau 3}=\frac{1}{\sqrt{2}}~.
\label{PMNSmatrixelemnts}
\end{align}
Therefore, the mixing angles  are expressed as follows:
\begin{align}
\sin ^2\theta _{12}&=\frac{|U_{e2}|^2}{|U_{e1}|^2+|U_{e2}|^2} \nonumber \\
&=\frac{1}{6(3+\cos 2\lambda )}\bigg [9+3\cos 2\lambda (1+\sin ^2\eta )+\sin ^2\eta (1+8\cos \psi _\ell \sin 2\lambda ) \nonumber \\
&-\cos ^2\eta (1+3\cos 2\lambda +8\cos \psi _\ell \sin 2\lambda ) \nonumber \\
&+2\sqrt{2}\sin 2\eta \Big \{ \cos \psi (1+3\cos 2\lambda -\cos \psi _\ell \sin 2\lambda )-3\sin \psi \sin \psi _\ell \sin 2\lambda \Big \} \bigg ]~, \nonumber \\
\sin ^2\theta _{23}&=\frac{|U_{\mu 3}|^2}{|U_{\mu 3}|^2+|U_{\tau 3}|^2}
=1-\frac{1}{2-\sin ^2\lambda }~,\qquad \sin ^2\theta _{13}=|U_{e3}|^2=\frac{\sin ^2\lambda }{2}~.
\label{lepton-mixing-angles}
\end{align}
In order to compare our model with the TBM in the neutrino sector,
it is useful to write down the mixing angles without the contribution from the charged lepton sector,  that is  ones in Eq. (\ref{neutrino-mixing-matrix}), 
\begin{align}
\sin ^2\theta ^\nu _{12}=\frac{1}{3}\left (1+\sin ^2\eta +\sqrt{2}\sin 2\eta \cos \psi \right )~,
\qquad 
\sin ^2\theta _{23}^\nu=\frac{1}{2} \ , \qquad \sin\theta _{13}^\nu=0 \ ,
\label{neutrinoside1}
\end{align}
where $\theta_{ij}^\nu$ denote the mixing angles only in the neutrino sector.
Our model is different from the TBM only in $\theta_{12}^\nu$.
We will see how large our $\theta_{12}^\nu$ is deviated from the TBM $\sin^2\theta_{12}^\nu=1/3$ in the next section.

Before closing this subsection, we present a simple sum rule between  $\theta _{23}$ and 
 $\theta _{13}$  in Eq.(\ref{lepton-mixing-angles}):
\begin{equation}
\sin ^2\theta _{23}=1-\frac{1}{2\cos ^2\theta _{13}}~.
\label{sumrule}
\end{equation}
By using the experimental data~\cite{Gonzalez-Garcia:2014bfa} for $3\sigma $ range, 
the $\sin ^2\theta _{23}$ is predicted to be  
\begin{equation}
0.487\leq \sin ^2\theta _{23}\leq 0.490,
\end{equation}
which is within the $3\sigma $ range for the experimental data~\cite{Gonzalez-Garcia:2014bfa} as $0.389\leq \sin ^2\theta _{23}\leq 0.664$. 
We show the details of our numerical analyses in section~\ref{sec:numerical}.

\subsection{$S_4$ flavor model for $m_3\not =0$}
We can obtain  another  simple model with  the non-vanishing  lightest neutrino mass. 
The particle assignments are same as the model in Table~\ref{tab:assignment}. 
Therefore, the $S_4\times Z_4$ invariant Lagrangian for the Yukawa coupling  is same as 
in Eq.~(\ref{lagrangian}).
The VEV alignments for the neutrino sector are different from Eq.~(\ref{alignments}). 
Taking VEV alignments as 
\footnote{The alignments $\langle \phi _\nu \rangle /v_\nu =(0,1,1),~(0,-1,-1),~(0,-1,1)$ also 
lead to the same $M_\nu$ in Eq.~(\ref{model2}).}
\begin{equation}
\langle \phi _\nu \rangle = v_\nu (0,1,-1)\ ,\qquad \langle \phi _\nu '\rangle = v_\nu '(1,1,1)\ ,
\end{equation}
we obtain the Dirac neutrino mass matrix $M_D$ as
\begin{equation}
M_D=
\begin{pmatrix}
0 & 0 & y_D'v_\nu ' \\
\frac{1}{\sqrt{2}}y_Dv_\nu & \frac{1}{\sqrt{6}}y_Dv_\nu & y_D'v_\nu ' \\
\frac{1}{\sqrt{2}}y_Dv_\nu & -\frac{1}{\sqrt{6}}y_Dv_\nu & y_D'v_\nu '
\end{pmatrix}\frac{v}{\Lambda }\ .
\end{equation}
The right-handed Majorana neutrino mass matrix $M_N$ is same in Eq.~(\ref{Majoranamatrix}) as 
\begin{equation}
M_N=
\begin{pmatrix}
M_1 & 0 & 0 \\
0 & M_1 & 0 \\
0 & 0 & M_2
\end{pmatrix}\ .
\end{equation}
By using the seesaw mechanism, the left-handed Majorana neutrino mass matrix $M_\nu $ is written as
\begin{equation}
M_\nu =
\begin{pmatrix}
b & b & b \\
b & 2a+b & a+b \\
b & a+b & 2a+b
\end{pmatrix}, \qquad a=\frac{(y_Dv_\nu v)^2}{3M_1\Lambda ^2},\qquad b=\frac{(y_D'v_\nu 'v)^2}{M_2\Lambda ^2}\ ,
\label{model2}
\end{equation}
where $a$ and $b$ are complex parameters. 
Moving to the  TBM base, the  neutrino mass matrix is given  as
\begin{equation}
M_\nu =V_\text{TBM}
\begin{pmatrix}
a & -\sqrt{2}a & 0 \\
-\sqrt{2}a & 2a+3b & 0 \\
0 & 0 & a
\end{pmatrix}V_\text{TBM}^T\ .
\end{equation}
In order to get the mixing angles of the left-handed Majorana neutrino, we investigate $M_\nu ^\dagger M_\nu $ as 
\begin{equation}
M_\nu ^\dagger M_\nu =V_\text{TBM}
\begin{pmatrix}
3|a|^2 & -3\sqrt{2}(|a|^2+|a||b|e^{-i\varphi }) & 0 \\
-3\sqrt{2}(|a|^2+|a||b|e^{i\varphi }) & 3(2|a|^2+3|b|^2+4|a||b|\cos \varphi ) & 0 \\
0 & 0 & |a|^2
\end{pmatrix}V_\text{TBM}^T \ ,
\end{equation}
where $a=|a|e^{i\varphi _a}$, $b=|b|e^{i\varphi _b}$, and $\varphi _a-\varphi _b\equiv \varphi $~. 
Hereafter, we define $|a|\equiv a$ and $|b|\equiv b$ for simplicity. 
Then, the neutrino mass eigenvalues are 
\begin{align}
m_1^2&=\frac{3}{2}\left [3a^2+3b^2+4ab\cos \varphi -\sqrt{\left (3a^2+3 b^2+4ab\cos \varphi \right )^2-4a^2b^2}\right ], \nonumber \\
m_2^2&=\frac{3}{2}\left [3a^2+3b^2+4ab\cos \varphi +\sqrt{\left (3a^2+3 b^2+4ab\cos \varphi \right )^2-4a^2b^2}\right ], \nonumber \\
m_3^2&=a^2.
\end{align}
Therefore, the neutrino mass squared differences are 
\begin{align}
\Delta m_{\text{atm}}^2&=m_2^2-m_3^2=\frac{3}{2}\left [\frac{7}{3}a^2+3b^2+4ab\cos \varphi +\sqrt{\left (3a^2+3b^2+4ab\cos \varphi \right )^2-4a^2b^2}\right ]~, \nonumber \\
\Delta m_{\text{sol}}^2&=m_2^2-m_1^2=3\sqrt{\left (3a^2+3b^2+4ab\cos \varphi \right )^2-4a^2b^2}~,
\end{align}
and the neutrino mixing matrix $U_\nu $ is 
\begin{equation}
U_\nu =V_\text{TBM}
\begin{pmatrix}
\cos \eta & e^{-i\psi }\sin \eta & 0 \\
-e^{i\psi }\sin \eta & \cos \eta & 0 \\
0 & 0 & 1
\end{pmatrix},
\end{equation}
where
\begin{equation}
\tan 2\eta =\frac{2a\sqrt{2(a^2+b^2+2ab\cos \varphi )}}{a^2+3b^2+4ab\cos \varphi }~, 
\qquad \tan \psi =-\frac{b\sin \varphi }{a+b\cos \varphi }~.
\label{nonzeromixing}
\end{equation}
We can easily find the mass ordering with $m_2 > m_1 > m_3$
for any $\cos \varphi$.
Therefore, the neutrino mass hierarchy is only inverted one 
and the neutrino mixing angles are determined by two neutrino mass squared differences and relative phase $\varphi$.

The charged lepton mass matrix is the same as the one in the previous subsection. Therefore, 
the PMNS mixing matrix elements are expressed in the same form in Eq.~(\ref{PMNSmatrixelemnts}).
We have the same sum rule between  $\theta _{23}$ and  $\theta _{13}$ in Eq.~(\ref{sumrule}).

\subsection{Potential analysis and VEV alignments}
\label{sec:potential}
In this subsection, we present the potential analysis of flavons in the framework of
the supersymmetry with the $U(1)_R$ symmetry.
We can generate the vacuum alignment through $F$-terms by coupling flavons to driving 
fields, which carry the $R$ charge $+2$ under $U(1)_R$ symmetry. We also assign $R$ charge $+1$ to the lepton doublets, right-handed charged leptons, and right-handed Majorana neutrinos. 
In addition to driving fields with $R=2$, we introduce 
 a new flavon $\phi _\nu ''$, which does not couple to leptons and right-handed Majorana neutrinos due to the $U(1)_{FN}$ charge. 
The charge assignments of the additional flavon and driving fields are summarized in Table~\ref{tab:assignment-additional}.
\begin{table}[hbtp]
\begin{center}
\begin{tabular}{|c|c||cccc|}
\hline 
& $\phi _\nu ''$ & $\xi _0$ & $\xi _0'$ & $\eta _0$ & $\chi _0$ \\
\hline 
$SU(2)$ & ${\bf 1}$ & ${\bf 1}$ & ${\bf 1}$ & ${\bf 1}$ & ${\bf 1}$ \\
$S_4$ & ${\bf 3}$ & ${\bf 1}$ & ${\bf 1'}$ & ${\bf 1}$ & ${\bf 2}$ \\
$Z_4$ & $-1$ & $1$ & $1$ & $i$ & $-1$ \\
$U(1)_{FN}$ & $-z$ & $\ell $ & $\ell $ & $0$ & $z$ \\
$U(1)_R$ & $0$ & $2$ & $2$ & $2$ & $2$ \\
\hline 
\end{tabular}
\caption{The assignments of the additional fields.}
\label{tab:assignment-additional}
\end{center}
\end{table}

Let us write down the $S_4\times Z_4\times U(1)_{FN}$ invariant superpotential for the  interaction of scalar fields as follow:
\begin{equation}
w_V=y_{\xi _0}\phi _\ell \phi _\ell ''\xi _0 +y_{\xi _0'}\phi _\ell '\phi _\ell ''\xi _0' +
y_{\eta _0} \phi _\nu \phi _\ell ''\eta _0 +y_{\chi _0}\phi _\nu '\phi _\nu ''\chi _0~,
\end{equation}
where $y_{\xi _0}$, $y_{\xi _0'}$, $y_{\eta _0}$ and  $y_{\chi _0}$ are arbitrary parameters.
Then, the potential $V$ is given as
\begin{align}
V=\left |\frac{\partial w_V}{\partial X_0}\right |^2&=\left |y_{\xi _0}(\phi _{\ell 1}\phi _{\ell 1}''+\phi _{\ell 2}\phi _{\ell 2}''+\phi _{\ell 3}\phi _{\ell 3}'')\right |^2
+\left |y_{\xi _0'}(\phi _{\ell 1}'\phi _{\ell 1}''+\phi _{\ell 2}'\phi _{\ell 2}''+\phi _{\ell 3}'\phi _{\ell 3}'')\right |^2 \nonumber \\
&+\left |y_{\eta _0}(\phi _{\nu 1}\phi _{\ell 1}''+\phi _{\nu 2}\phi _{\ell 2}''+\phi _{\nu 3}\phi _{\ell 3}'')\right |^2 \nonumber \\
&+\left |\frac{1}{\sqrt{2}}y_{\chi _0}(\phi _{\nu 2}'\phi _{\nu 2}''-\phi _{\nu 3}'\phi _{\nu 3}'')\right |^2
+\left |\frac{1}{\sqrt{6}}y_{\chi _0}(-2\phi _{\nu 1}'\phi _{\nu 1}''+\phi _{\nu 2}'\phi _{\nu 2}''+\phi _{\nu 3}'\phi _{\nu 3}'')\right |^2,
\label{eq:scalar-potential}
\end{align}
where $X_0=\xi _0, \xi _0',\eta _0,\chi _0$. 
Therefore, conditions to realize the potential minimum ($V=0$) are given from Eq.~(\ref{eq:scalar-potential}) as
\begin{align}
y_{\xi _0}(\phi _{\ell 1}\phi _{\ell 1}''+\phi _{\ell 2}\phi _{\ell 2}''+\phi _{\ell 3}\phi _{\ell 3}'')&=0, \nonumber \\
y_{\xi _0'}(\phi _{\ell 1}'\phi _{\ell 1}''+\phi _{\ell 2}'\phi _{\ell 2}''+\phi _{\ell 3}'\phi _{\ell 3}'')&=0, \nonumber \\
y_{\eta _0}(\phi _{\nu 1}\phi _{\ell 1}''+\phi _{\nu 2}\phi _{\ell 2}''+\phi _{\nu 3}\phi _{\ell 3}'')&=0, \nonumber \\
\frac{1}{\sqrt{2}}y_{\chi _0}(\phi _{\nu 2}'\phi _{\nu 2}''-\phi _{\nu 3}'\phi _{\nu 3}'')&=0, \nonumber \\
\frac{1}{\sqrt{6}}y_{\chi _0}(-2\phi _{\nu 1}'\phi _{\nu 1}''+\phi _{\nu 2}'\phi _{\nu 2}''+\phi _{\nu 3}'\phi _{\nu 3}'')&=0.
\end{align}
One of the solutions which satisfies these conditions is given as
\begin{align}
&\langle \phi _{\ell 3}\rangle =\langle \phi _{\ell 2}'\rangle =\langle \phi _{\ell 3}'\rangle 
=\langle \phi _{\ell 1}''\rangle =\langle \phi _{\ell 2}''\rangle =\langle \phi _{\nu 2}\rangle =\langle \phi _{\nu 3}\rangle =0, \nonumber \\
&\langle \phi _{\nu 1}'\rangle =\langle \phi _{\nu 2}'\rangle =\langle \phi _{\nu 3}'\rangle ,\quad 
\langle \phi _{\nu 1}''\rangle =\langle \phi _{\nu 2}''\rangle =\langle \phi _{\nu 3}''\rangle ,
\end{align}
then,  the VEV alignments are 
\begin{align}
&\langle \phi _\ell \rangle =(v_{\ell 1},v_{\ell _2},0),\quad \langle \phi _\ell '\rangle =v_\ell '(1,0,0),\quad \langle \phi _\ell ''\rangle =v_\ell ''(0,0,1), \nonumber \\
&\langle \phi _\nu \rangle =v_\nu (1,0,0),\quad \langle \phi _\nu '\rangle =v_\nu '(1,1,1),
\end{align}
which have been used in the subsection 2.1. We can also derive the VEV alignments of the subsection 2.2 
in the similar discussion.


\section{Numerical analyses}
\label{sec:numerical}
In this section, we show the numerical analyses in our $S_4$ models.
We use the result of the global analyses of neutrino oscillation experiments~\cite{Gonzalez-Garcia:2014bfa,Tortola:2012te,Fogli:2012ua}.
The $3\sigma $ range of the experimental data~\cite{Gonzalez-Garcia:2014bfa} for the inverted neutrino mass hierarchy are
\begin{align}
7.02&\leq \frac{\Delta m_{\rm sol}^2}{10^{-5}~\text{eV}^2}\leq 8.09, \qquad 2.307\leq 
\frac{|\Delta m_{\rm atm}^2|}{10^{-3}~\text{eV}^2}\leq 2.590, \nonumber \\
\nonumber \\
0.270\leq \sin ^2\theta _{12}&\leq 0.344, \quad 0.389\leq \sin ^2\theta _{23}\leq 0.664, \quad 0.0188\leq \sin ^2\theta _{13}\leq 0.0251.
\label{inputdata}
\end{align}

Our numerical strategy is as follows.
Fixing a random value for  $\varphi$ of Eq.~(\ref{mass-square-differences})
in the region  $-\pi \leq \varphi \leq \pi $, 
then, $a$ and $b$ are determined  from  the experimental data of  $\Delta m_{\text{atm}}^2$ and $\Delta m_{\text{sol}}^2$~.
Then, we can calculate the neutrino mixing matrix ($\eta $ and $\psi $) 
from Eqs.~(\ref{neutrino-mixing-matrix}) and (\ref{tan2eta-tanpsi}). 
The magnitude of  $\lambda $ is determined in $\sin^2 \theta_{13}$  of Eq.~(\ref{lepton-mixing-angles})
by using the experimental data~\cite{Gonzalez-Garcia:2014bfa} for $3\sigma $ range.
Finally, choosing  a random value for  $\psi_\ell$ in $-\pi \leq \psi _\ell \leq \pi$, 
we calculate the PMNS mixing matrix elements in Eq.~(\ref{PMNS-matrix}),
which are adjusted to the experimental data of Eq.(\ref{inputdata}).

In our numerical studies, we predict the Dirac phase and the Majorana phases. 
Our predicted PMNS matrix in the previous section should be compared with the following
 conventional parametrization \cite{PDG} including the Majorana phases:
\begin{footnotesize}
\begin{align}
&\begin{pmatrix}
e^{i\delta _e} & 0 & 0 \\
0 & e^{i\delta _\mu } & 0 \\
0 & 0 & e^{i\delta _\tau }
\end{pmatrix}
\begin{pmatrix}
c_{12} c_{13} & s_{12} c_{13} & s_{13}e^{-i\delta _{CP}} \\
-s_{12} c_{23} - c_{12} s_{23} s_{13}e^{i\delta _{CP}} & 
c_{12} c_{23} - s_{12} s_{23} s_{13}e^{i\delta _{CP}} & s_{23} c_{13} \\
s_{12} s_{23} - c_{12} c_{23} s_{13}e^{i\delta _{CP}} & 
-c_{12} s_{23} - s_{12} c_{23} s_{13}e^{i\delta _{CP}} & c_{23} c_{13}
\end{pmatrix}
\begin{pmatrix}
e^{i\alpha } & 0 & 0 \\
0 & e^{i\beta } & 0 \\
0 & 0 & 1
\end{pmatrix} \nonumber \\
&=
\begin{pmatrix}
c_{12} c_{13}e^{i(\delta _e-\alpha )} & s_{12} c_{13}e^{i(\delta _e-\beta )} & s_{13}e^{i(\delta _e-\delta _{CP})} \\
(-s_{12} c_{23} - c_{12} s_{23} s_{13}e^{i\delta _{CP}})e^{i(\delta _\mu -\alpha )} & 
(c_{12} c_{23} - s_{12} s_{23} s_{13}e^{i\delta _{CP}})e^{i(\delta _\mu -\beta )} & s_{23} c_{13}e^{i\delta _\mu } \\
(s_{12} s_{23} - c_{12} c_{23} s_{13}e^{i\delta _{CP}})e^{i(\delta _\tau -\alpha )} & 
(-c_{12} s_{23} - s_{12} c_{23} s_{13}e^{i\delta _{CP}})e^{i(\delta _\tau -\beta )} & c_{23} c_{13}e^{i\delta _\tau }
\end{pmatrix}~,
\label{PMNSmatrix}
\end{align}
\end{footnotesize}
where $c_{ij}$ and $s_{ij}$ denote $\cos \theta _{ij}$ and $\sin \theta _{ij}$, respectively.
The phases $\delta _e$, $\delta _\mu $, and $\delta _\tau $ could be
absorbed  in the left-handed charged lepton fields, $\delta_{CP}$ is the Dirac phase,
and  $\alpha $, $\beta$ are the Majorana phases. 


We can calculate the Dirac phase $\delta_{CP}$ as follows.
The Jarlskog invariant \cite{Jarlskog:1985ht},
 which is the measure describing the size of the CP violation, is given as 
\begin{align}
J_{CP}&=\text{Im}\left [U_{xi}U_{yj}U_{yi}^*U_{xj}^*\right ] \nonumber \\
&=\frac{\sin 2\lambda }{48}\left [3\sqrt{2}\cos \psi _\ell \sin \psi \sin 2\eta 
-\sin \psi _\ell \left (\sqrt{2}\cos \psi \sin 2\eta +4\cos 2\eta \right )\right ],
\end{align}
where $x,~y=e,~\mu ,~\tau $ and $i,~j=1$-$3$. Then, the CP violating Dirac phase is written 
in terms of the lepton mixing angles and $J_{\text{CP}}$ as 
\begin{align}
\sin \delta _{\text{CP}}&=\frac{J_{CP}}{s_{23}c_{23}s_{12}c_{12}s_{13}c_{13}^2}\ .
\end{align}
In our model, the one of the PMNS mixing matrix elements $U_{\tau 1}$ is written  as
\begin{equation}
|U_{\tau 1}|=\left |-\frac{\cos \eta }{\sqrt{6}}-\frac{e^{i\psi }\sin \eta }{\sqrt{3}}\right |
=\left |s_{12}s_{23}-c_{12}c_{23}s_{13}e^{i\delta _{\text{CP}}}\right |~,
\end{equation}
then, the CP violating Dirac phase is also written as 
\begin{equation}
\cos \delta _{\text{CP}}=\frac{s_{12}^2s_{23}^2+c_{12}^2c_{23}^2s_{13}^2-
\frac{1}{6}\cos^2\eta -\frac{1}{3}\sin ^2\eta -\frac{1}{3\sqrt{2}}\sin 2\eta \cos \psi}
{2s_{12}s_{23}c_{12}c_{23}s_{13}}~.
\end{equation}
Thus, $\delta_{CP}$ is determined up to the quadrant.

Next, let us calculate  the Majorana phases.
In our model, $U_{\mu 3}$ and $U_{\tau 3}$ have no phases as seen in Eq. (\ref{PMNSmatrixelemnts}), so we find   $\delta_\mu =\delta_\tau =0$ in Eq. (\ref{PMNSmatrix}).
For convenience, expressing the phases of the mixing matrix  of  Eq. (\ref{PMNSmatrixelemnts}) as
\begin{equation}
U_{e1}=|U_{e1}|e^{i\delta _1},\qquad U_{e2}=|U_{e2}|e^{i\delta _2},
\qquad U_{e3}=|U_{e3}|e^{-i\psi_\ell},
\end{equation}
 we derive the following relations by comparing  with  Eq. (\ref{PMNSmatrix}):
\begin{equation}
\delta_1=\delta _e-\alpha \ ,\qquad \delta _2=\delta _e-\beta \  ,\qquad
 \psi _\ell=\delta _{CP}-\delta _e \ .
\end{equation}
 Eliminating $\delta _e$ in these relations,  we obtain 
\begin{equation}
\alpha =\delta _{CP}-\psi _\ell -\delta _1,\qquad \beta =\delta _{CP}-\psi _\ell -\delta _2,
\qquad \alpha -\beta =\delta _2-\delta _1,
\label{Majorana-phase}
\end{equation}
from which we can calculate the Majorana phases numerically.

\subsection{$S_4$ flavor  model for $m_3=0$}
\begin{wrapfigure}{r}{8cm}
\vspace{-5mm}
\includegraphics[width=7.5cm]{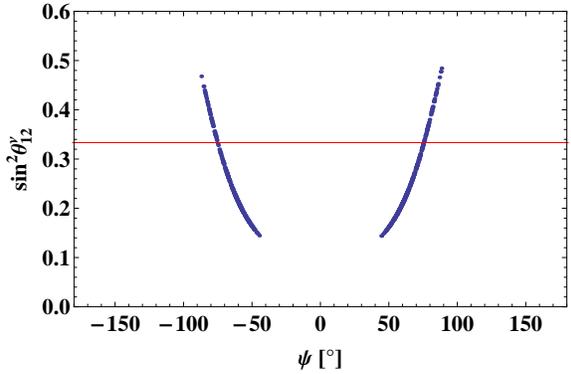}
\caption{The allowed region on $\psi$--$\sin^2\theta_{12}^\nu$ plane.
The red line corresponds to the TBM. }
\label{fig:tribi1}
\end{wrapfigure}
In this subsection, we show the numerical results in the case of the lightest neutrino mass $m_3=0$.
In order to reproduce the relevant mass squared differences, $\Delta m_{\rm atm}^2$
and $\Delta m_{\rm sol}^2$ for the inverted mass hierarchy, 
 $a\simeq 3b$ and $\varphi \simeq \pm \pi$ should be  satisfied. 
Then, the $\psi$ of Eq. (\ref{tan2eta-tanpsi}) is restricted 
in $45^\circ\lesssim |\psi| \lesssim 90^\circ$.
On the other hand, $\sin\eta$ in Eq. (\ref{tan2eta-tanpsi})
is restricted  in $-0.70\lesssim \sin\eta \lesssim -0.38$.
In Fig.\ref{fig:tribi1}, we show the allowed region on $\psi$--$\sin^2\theta_{12}^\nu$ plane, where
$\theta_{12}^\nu$ is the mixing angle of the neutrino sector 
without the charged lepton contribution as seen in Eq.(\ref{neutrinoside1}).
As seen in this figure, our model is possibly deviated from the TBM considerably.
Then, the mixing angle and the phase of the charged lepton sector become important
to adjust the observed PMNS mixing angles. 

\begin{figure}[h!]
\begin{minipage}[]{0.45\linewidth}
\includegraphics[width=7.5cm]{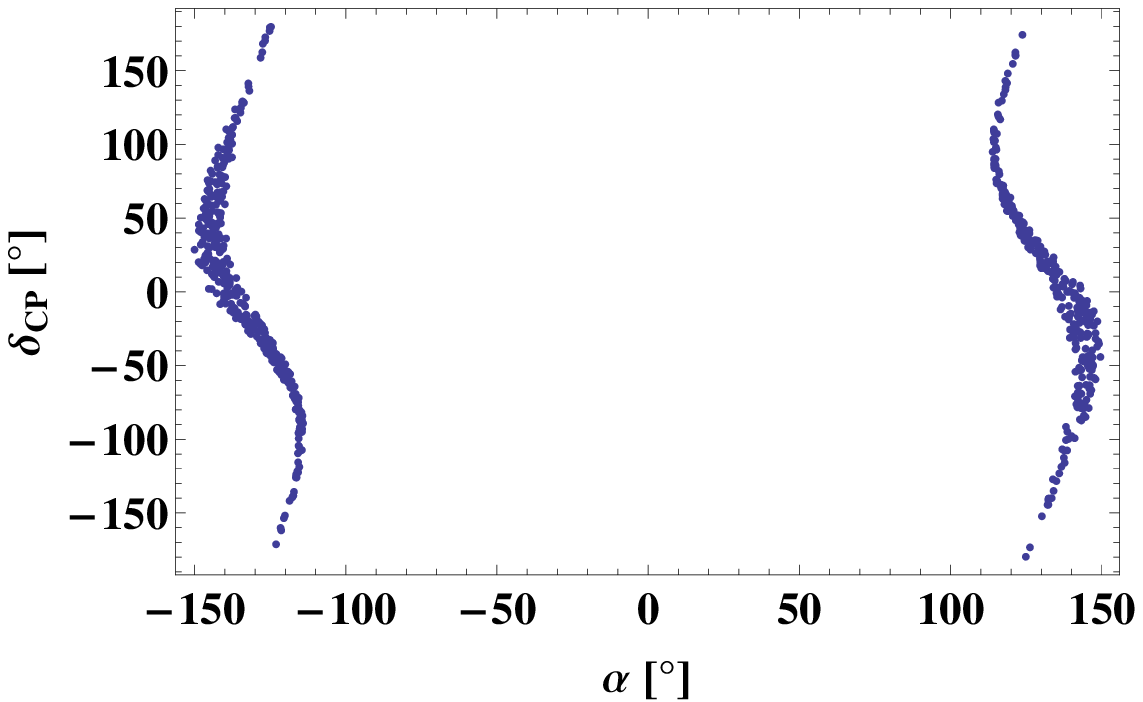}
\caption{The predicted region on $\alpha $--$\delta _{CP}$ plane. }
\label{fig:alphadelta}
\end{minipage}
\hspace{5mm}
\begin{minipage}[]{0.45\linewidth}
\includegraphics[width=7.5cm]{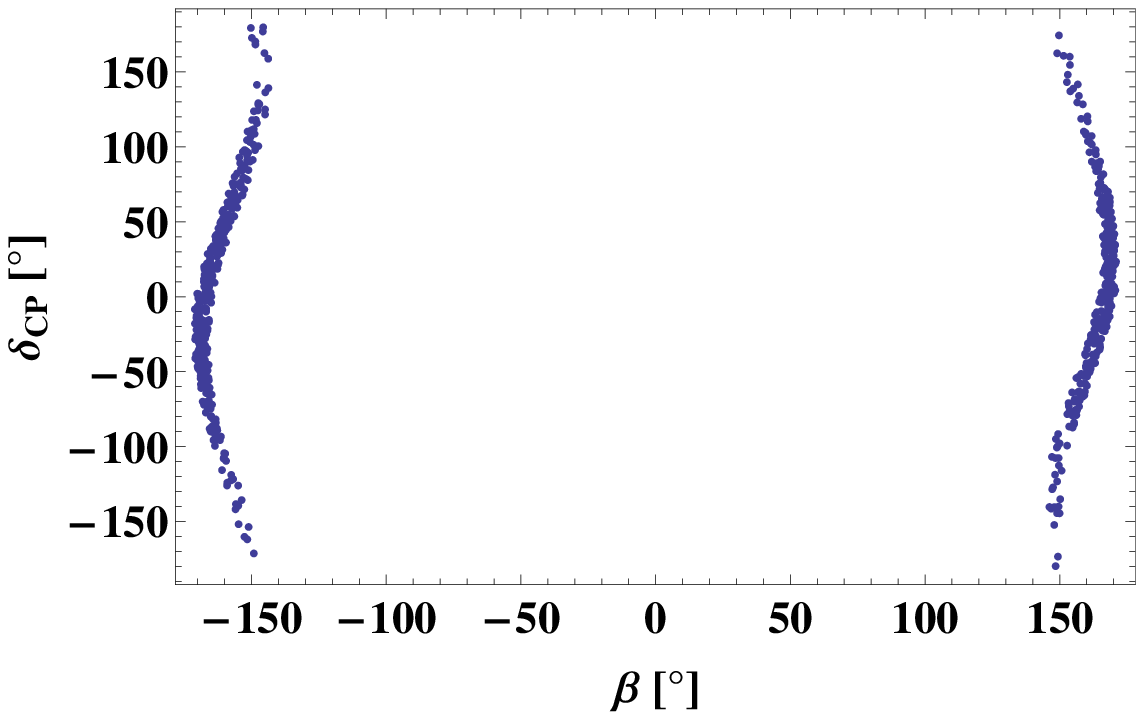}
\caption{The predicted region on  $\beta $--$\delta _{CP}$ plane. 
}
\label{fig:betadelta}
\end{minipage}
\end{figure}

Now we can calculate the Dirac phase and the Majorana phases.
In Figs.~\ref{fig:alphadelta} and \ref{fig:betadelta}, we show the  Dirac phase $\delta _{\text{CP}}$ versus the Majorana phases $\alpha $ and $\beta $, respectively.
The both positive and negative signs are allowed for $\alpha$ and $\beta$. 
The allowed regions of the Majorana phases are 
$110 ^\circ \lesssim |\alpha |\lesssim 150 ^\circ $ 
and $145 ^\circ \lesssim |\beta |\lesssim 170 ^\circ $.
On the other hand, the Dirac phase is still allowed 
in the all region as $-\pi\lesssim \delta_{CP}\lesssim \pi$. 

\begin{figure}[h!]
\begin{minipage}[]{0.45\linewidth}
\vspace{-4mm}
\includegraphics[width=7.5cm]{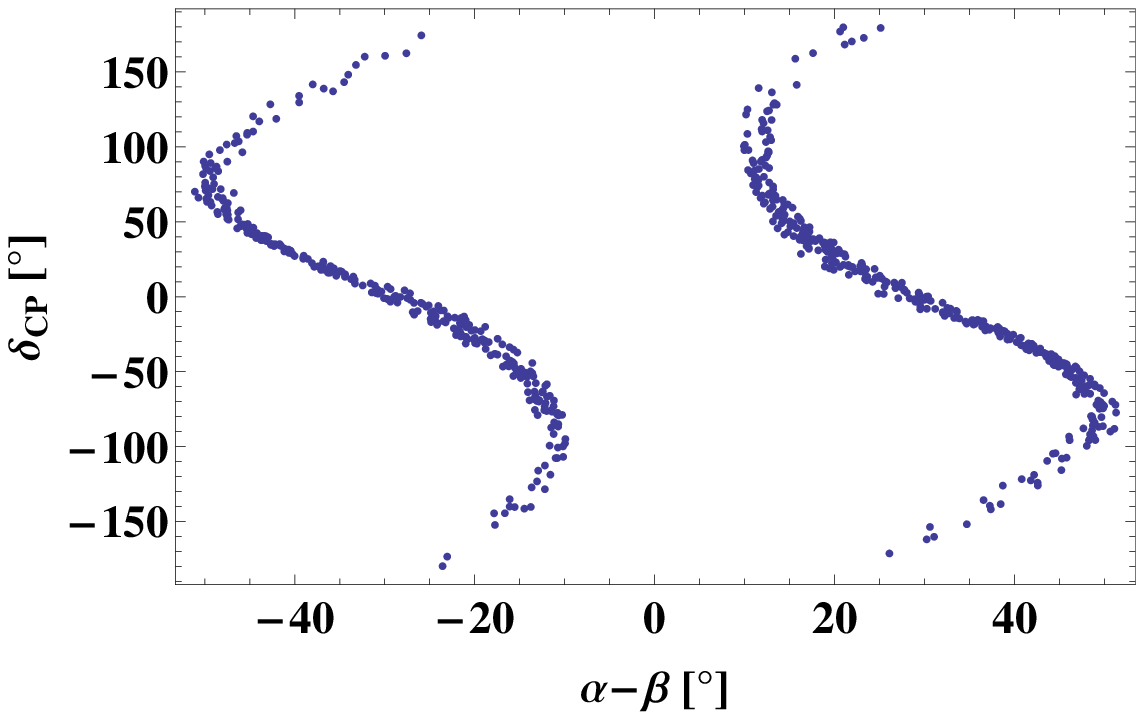}
\caption{The predicted region on $(\alpha -\beta )$--$\delta _{CP}$ plane. }
\label{fig:alphabetadelta}
\end{minipage}
\hspace{5mm}
\begin{minipage}[]{0.45\linewidth}
\includegraphics[width=7.5cm]{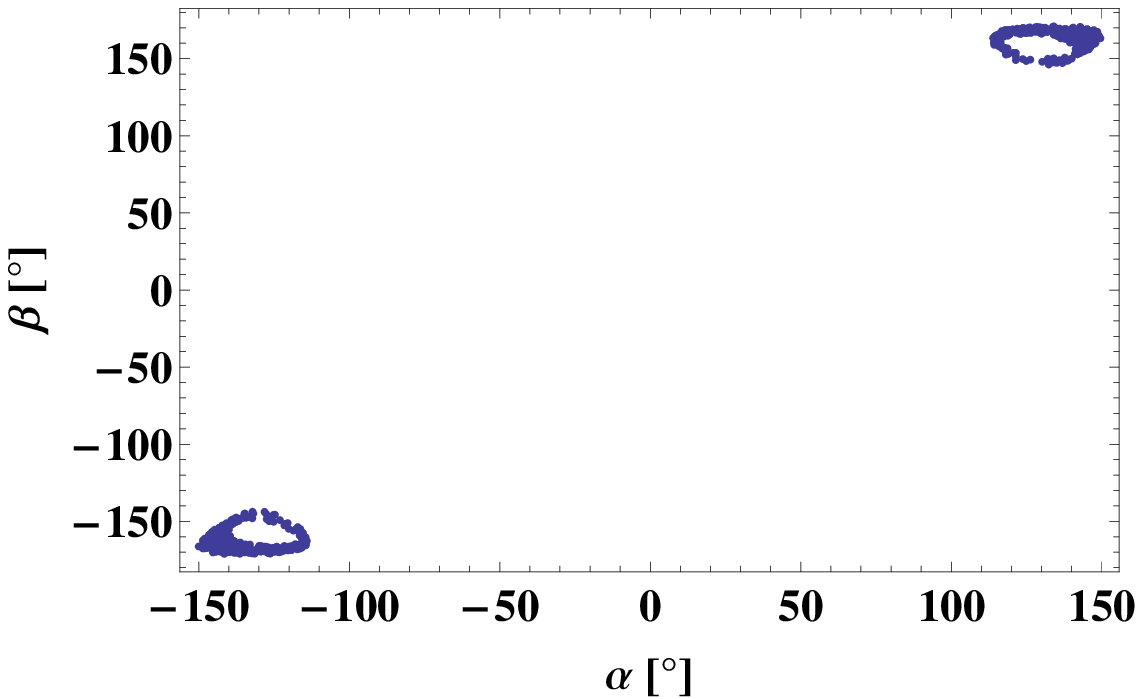}
\caption{The predicted region on $\alpha $--$\beta $ plane. The Majorana phases are restricted in the conner on this plane.}
\label{fig:alphabeta}
\end{minipage}
\end{figure}

Since the difference between the Majorana phases $\alpha$ and $\beta$ 
contributes to the $0\nu \beta \beta $, 
we show the Dirac phase versus $(\alpha-\beta )$ in Fig.~\ref{fig:alphabetadelta}.
We obtain  $10^\circ\lesssim |\alpha -\beta |\lesssim 50^\circ$. 
We also show the allowed region of the Majorana phases 
on the $\alpha$--$\beta $ plane in Fig.~\ref{fig:alphabeta}. 
The Majorana phases $\alpha$ and $\beta$ are restricted in the two regions on this plane.


\begin{figure}[h!]
\begin{minipage}[]{0.45\linewidth}
\includegraphics[width=7.5cm]{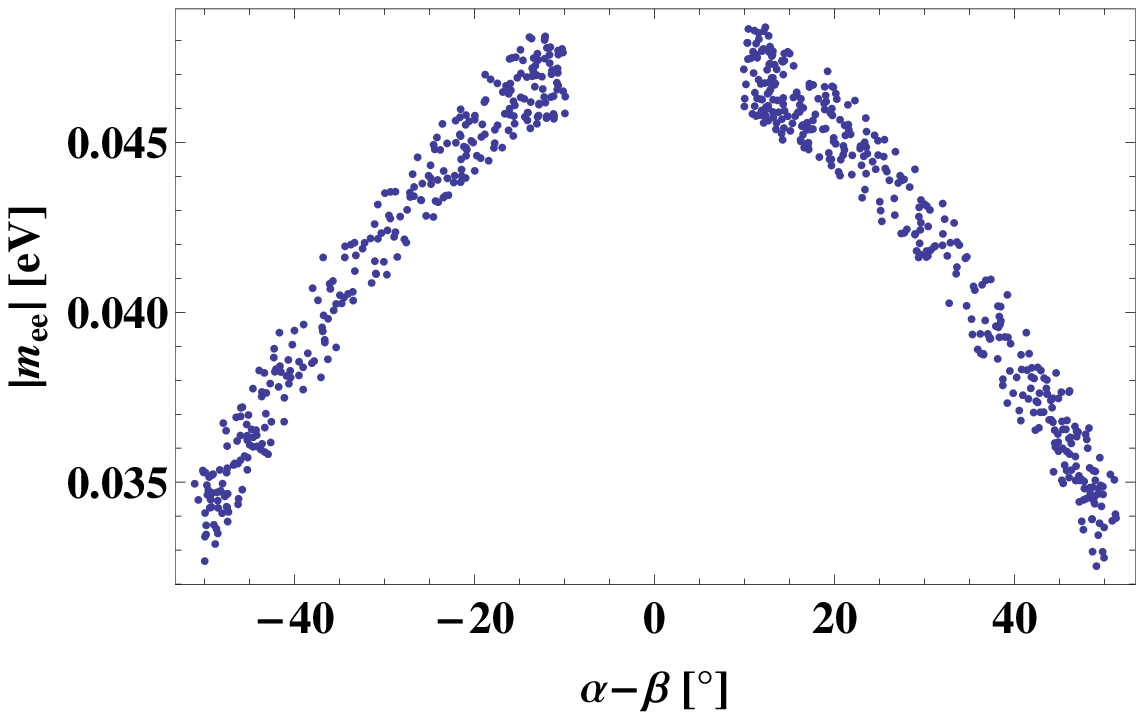}
\caption{The predicted  effective neutrino mass for the $0\nu \beta \beta $  versus $(\alpha -\beta )$.}
\label{fig:alphabetamee}
\end{minipage}
\hspace{5mm}
\begin{minipage}[]{0.45\linewidth}
\includegraphics[width=7.5cm]{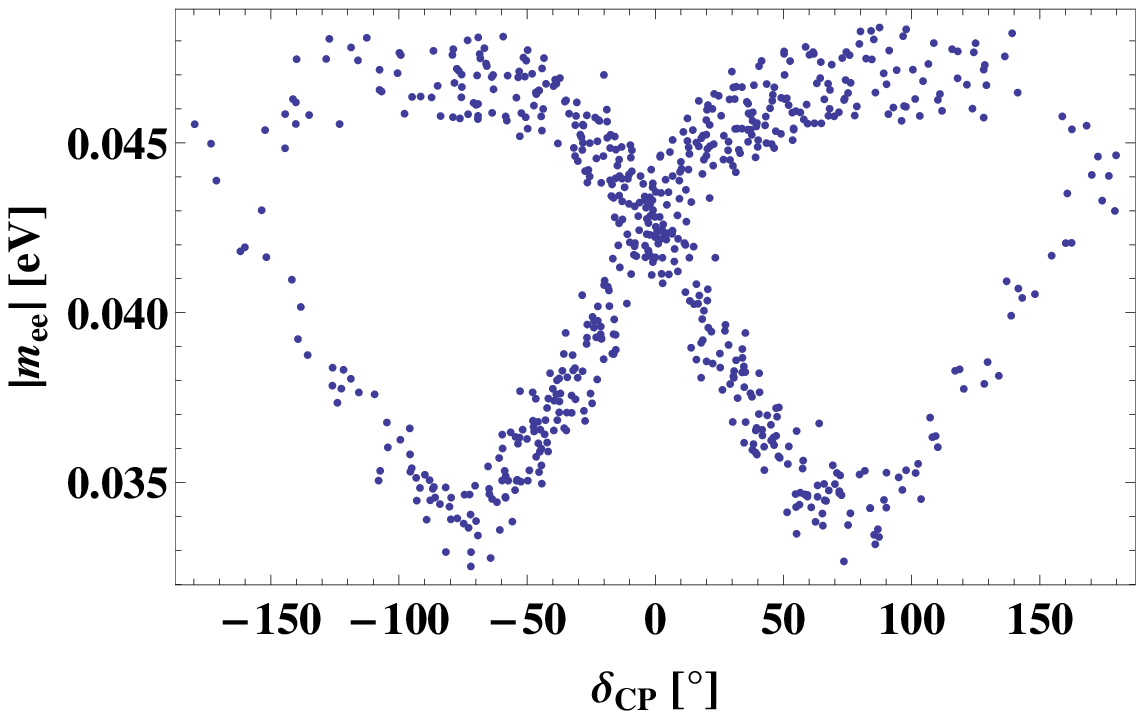}
\caption{The predicted effective neutrino mass for the $0\nu \beta \beta $  versus $\delta_{CP}$.}
\label{fig:deltamee}
\end{minipage}
\end{figure}

Based on the numerical results of the phases, we can estimate the effective neutrino mass for 
the $0\nu \beta \beta $.
The effective neutrino mass is written as 
\begin{equation}
m_{ee} \equiv \sum _{i=1}^3m_iU_{ei}^2~.
\end{equation}
We show the effective neutrino mass of the $0\nu \beta \beta $ versus the difference of the  Majorana phases and the  Dirac phase for  in Figs.~\ref{fig:alphabetamee} and
 \ref{fig:deltamee}, respectively.
 For the fixed effective neutrino mass for the $0\nu \beta \beta $, there are 
two fold degeneracy and four fold degeneracy
 in  the Majorana phase $(\alpha-\beta)$ and  the Dirac phase $\delta_{CP}$, respectively. 
 Therefore, the degeneracy will be solved if both $\delta_{CP}$ and $m_{ee}$ are observed.
 At present, we predict
$32~\text{meV}\lesssim |m_{ee}|\lesssim 49~\text{meV}$,
which is  close to the expected reaches
of the coming experiments of the $0\nu \beta \beta $ \cite{Gomez-Cadenas:2015twa}.

Finally, we show the sum of the neutrino masses is predicted as 
\begin{equation}
 0.0952~\text{eV}\lesssim {\displaystyle \sum_i}m_i\lesssim 0.101~\text{eV}.
 \end{equation}
This total neutrino mass is 
within the reaches of the future cosmological and astrophysical measurements,
such as the CMB spectrum, the galaxy distributions, and the red-shifted 21cm line
in the future \cite{Abazajian:2013oma}.

\subsection{$S_4$ flavor model for $m_3\not =0$}
\begin{wrapfigure}{r}{8cm}
\vspace{-9mm}
\includegraphics[width=7.5cm]{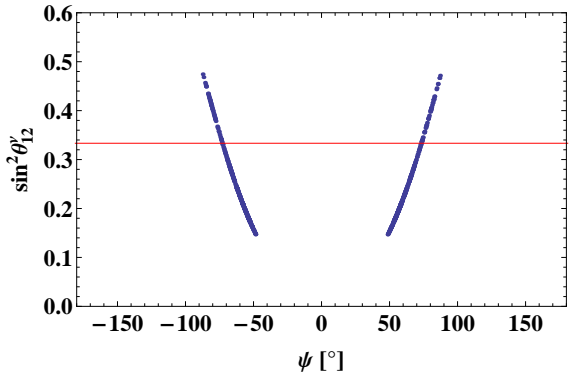}
\caption{The allowed region on $\psi$--$\sin^2\theta_{12}^\nu$ plane.
The red line corresponds to the TBM.}
\label{fig:tribi2}
\end{wrapfigure}
In this subsection, we show the numerical results in the case of the non-vanishing 
lightest neutrino mass $m_3\not=0$.
In order to reproduce the relevant mass squared differences, $\Delta m_{\rm atm}^2$
and $\Delta m_{\rm sol}^2$ for the inverted mass hierarchy, 
 $a\simeq b$ and $\varphi \simeq \pm \pi$ are satisfied. 
Then, the $\psi$ of Eq. (\ref{nonzeromixing}) is restricted  in 
$50^\circ\lesssim |\psi| \lesssim 90^\circ$ numerically.
On the other hand, $\sin\eta$ in Eq. (\ref{nonzeromixing})
is restricted  in $-0.70\lesssim \sin\eta \lesssim -0.54$.

In Fig. \ref{fig:tribi2}, we show the allowed region on $\psi$--$\sin^2\theta_{12}^\nu$ plane,
 where $\theta_{12}^\nu$ is the mixing angle of the neutrino sector 
without the charged lepton contribution as seen in Eq.(\ref{neutrinoside1}).
This result is very similar to the case of $m_3=0$ of Fig. \ref{fig:tribi1} 
in the previous subsection.

\begin{figure}[h!]
\begin{minipage}[]{0.45\linewidth}
\includegraphics[width=7.5cm]{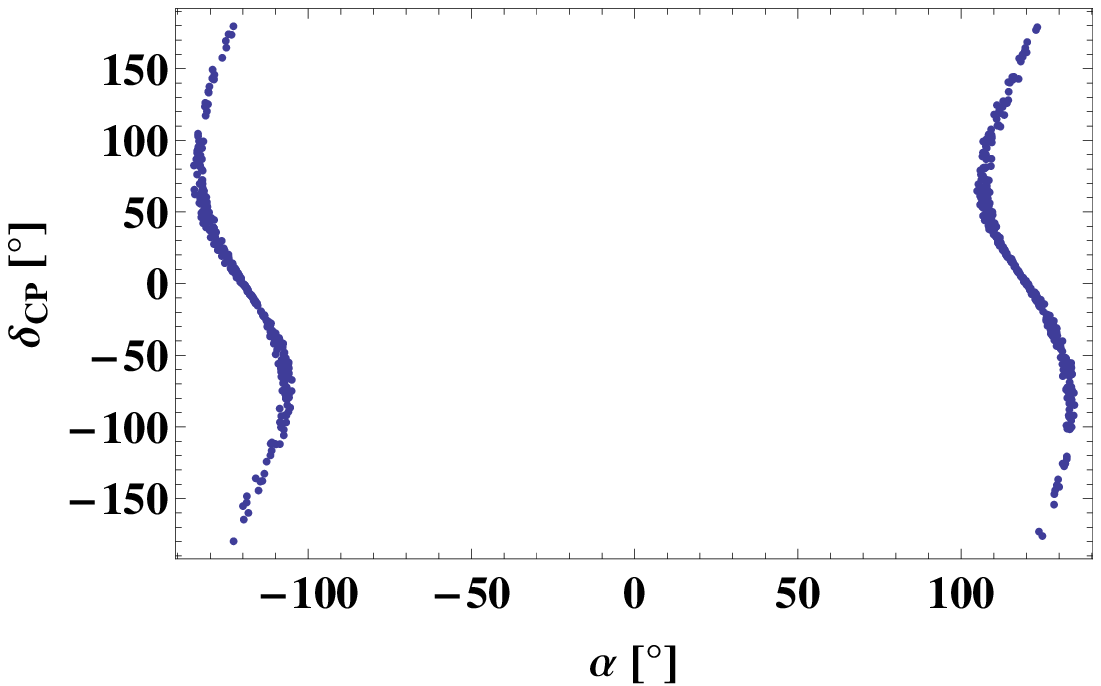}
\caption{The predicted region  on $\alpha $--$\delta _{CP}$ plane. }
\label{fig:otheralphadelta}
\end{minipage}
\hspace{5mm}
\begin{minipage}[]{0.45\linewidth}
\includegraphics[width=7.5cm]{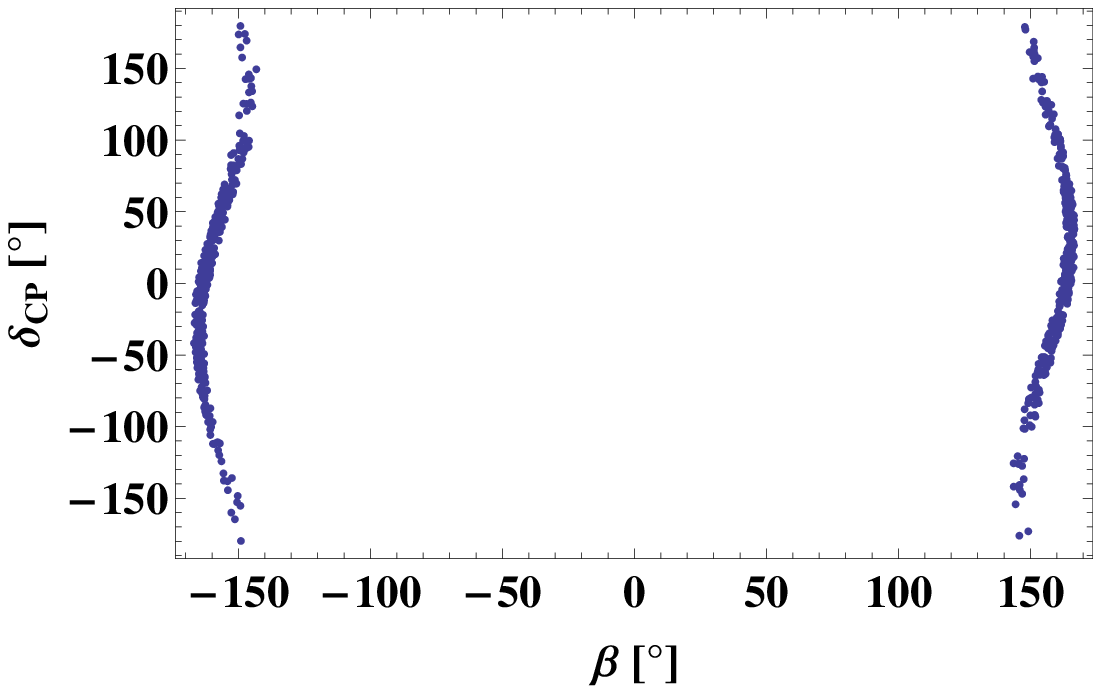}
\caption{The predicted region on $\beta $--$\delta _{CP}$ plane. }
\label{fig:otherbetadelta}
\end{minipage}
\end{figure}
\begin{figure}[h!]
\begin{minipage}[]{0.45\linewidth}
\includegraphics[width=7.5cm]{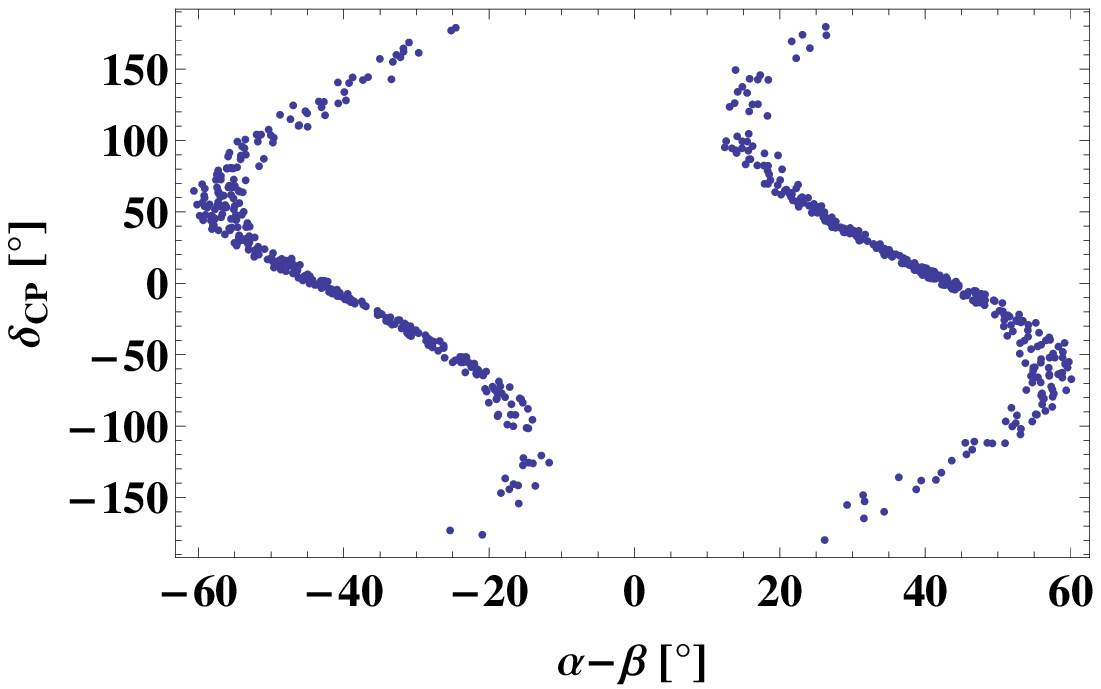}
\caption{The predicted region on $(\alpha -\beta )$--$\delta _{CP}$ plane. }
\label{fig:otheralphabetadelta}
\end{minipage}
\hspace{5mm}
\begin{minipage}[]{0.45\linewidth}
\includegraphics[width=7.5cm]{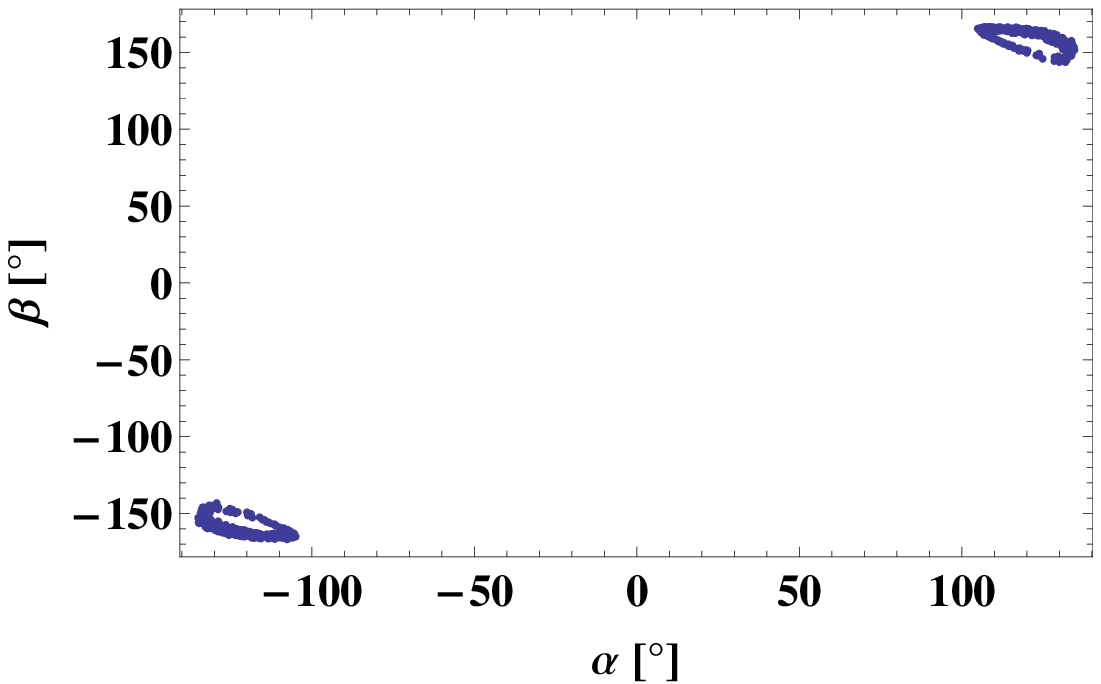}
\caption{The predicted  region on $\alpha $--$\beta$ plane. The Majorana phases are restricted in the conner of this plane.}
\label{fig:otheralphabeta}
\end{minipage}
\end{figure}
\newpage 

At first, in Figs.~\ref{fig:otheralphadelta} and \ref{fig:otherbetadelta}, 
we show the  Dirac phase $\delta_{\text{CP}}$ versus the Majorana phases $\alpha $ and $\beta $, respectively.
The both positive and negative signs are allowed for $\alpha $ and $\beta $. 
The allowed regions of the Majorana phases are 
$105 ^\circ \lesssim |\alpha |\lesssim 135 ^\circ $ 
and $145 ^\circ \lesssim |\beta |\lesssim 170 ^\circ $. 
The Dirac phase is still allowed in the all region as $-\pi\lesssim \delta_{CP}\lesssim \pi$.
We also show the Dirac phase $\delta _{CP}$ versus $(\alpha -\beta )$  
in Fig.~\ref{fig:otheralphabetadelta}.
We obtain $10^\circ\lesssim |\alpha -\beta |\lesssim 60^\circ$. 
In Fig.~\ref{fig:otheralphabeta}, we show the allowed region of the Majorana phases 
$\alpha$ and $\beta$.
The Majorana phases are restricted in the two regions on this plane as well as the case  of
the previous subsection.

\begin{figure}[h!]
\begin{minipage}[]{0.45\linewidth}
\includegraphics[width=7.5cm]{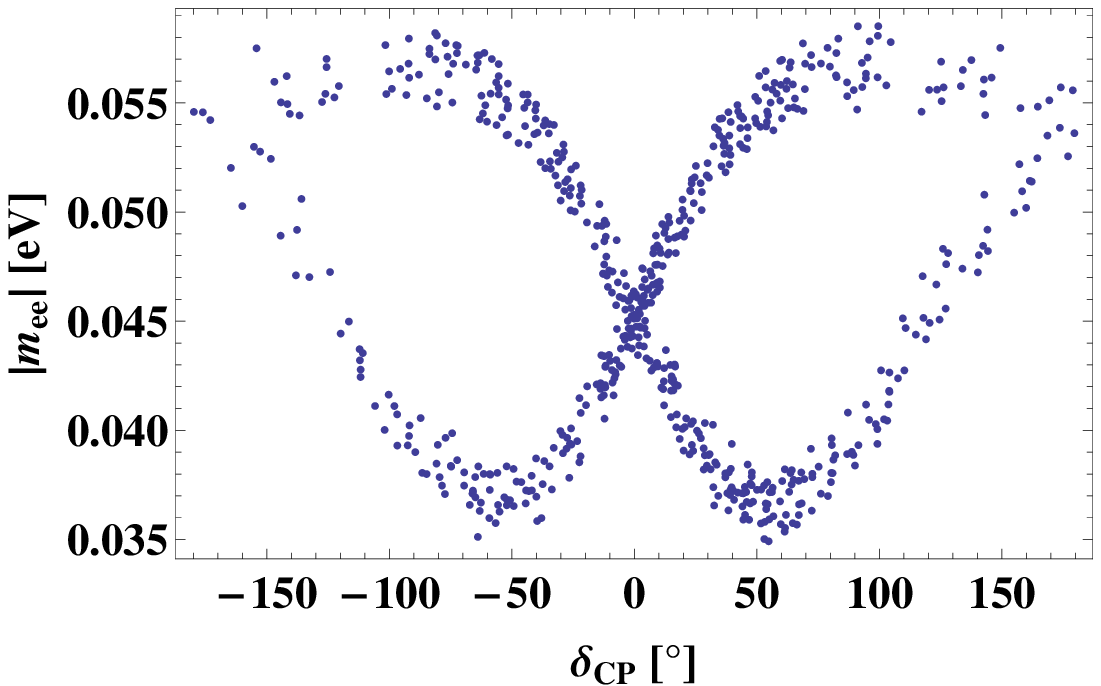}
\caption{The prediction on $\delta _{CP}$--$|m_{ee}|$ plane. 
The allowed region of effective neutrino mass for the $0\nu \beta \beta $ is $34~\text{meV}\lesssim |m_{ee}|\lesssim 59~\text{meV}$.}
\label{fig:otherdeltamee}
\end{minipage}
\hspace{5mm}
\begin{minipage}[]{0.45\linewidth}
\includegraphics[width=7.5cm]{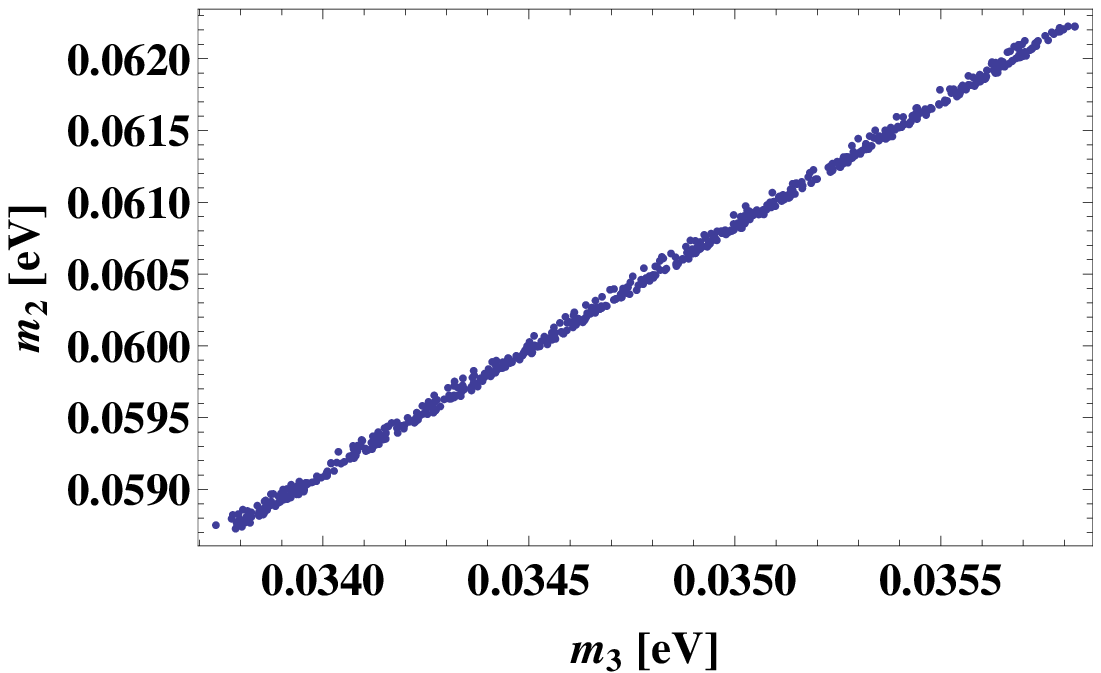}
\caption{The allowed region on $m_3$--$m_2$ plane. 
The lightest neutrino mass $m_3$ is proportional to  $m_2$.}
\label{fig:otherm3m2}
\end{minipage}
\end{figure}

Next, we also show the effective neutrino mass $m_{ee}$ for  the $0\nu \beta \beta $ versus 
the  Dirac phase 
  in Fig.~\ref{fig:otherdeltamee}.
 For the fixed effective neutrino mass for the $0\nu \beta \beta $, there are 
 four fold degeneracy in the Dirac phase $\delta_{CP}$ as well as the case of the previous subsection. 
 This degeneracy will be solved if both $\delta_{CP}$ and $m_{ee}$ are observed.
 The predicted magnitude of $m_{ee}$ is 
$34~\text{meV}\lesssim |m_{ee}|\lesssim 59~\text{meV}$,
which is  close to the expected reaches
of the coming experiments of the $0\nu \beta \beta $ \cite{Gomez-Cadenas:2015twa}.

Finally, we show the allowed region of neutrino masses $m_2$ and $m_3$  
in Fig.~\ref{fig:otherm3m2}. 
The neutrino mass $m_2$ is $58.7~\text{meV}\lesssim m_2\lesssim 62.2~\text{meV}$.
The lightest neutrino mass $m_3$ is $31.0~\text{meV}\lesssim m_3\lesssim 35.8~\text{meV}$.
The sum of the neutrino masses is predicted as
\begin{equation}
 0.150~\text{eV}\lesssim {\displaystyle \sum_i}m_i\lesssim 0.160~\text{eV}.
\end{equation}
This value is also  within the reaches of the future cosmological and astrophysical measurements.

\section{Discussions and Summary}
We have presented the minimal $S_4$ flavor models of leptons.
By using the alignments of the VEV's of the flavons obtained in subsection 2.3, 
the models lead to the two different neutrino mass spectra with the inverted mass hierarchy. 
The neutrino mass matrix is given by two complex parameters in Eqs.(\ref{model1}) and (\ref{model2}).
The first case has one vanishing neutrino mass $m_3=0$, and the second case
 has three non-vanishing neutrino masses.
The charged leptons are not flavor diagonal. 
There remains the first and second family mixing, which give the non-vanishing
 reactor angle $\theta_{13}$.
We have moved to the TBM basis of neutrinos 
 in order to clarify the deviation from the TBM.
 

It is not necessary to go to the TBM basis to calculate the mixing angles, 
but this is merely a convenient basis to do so. We present alternative basis,
 the bi-maximal base or the $\mu $-$\tau $ symmetric one, where it is also convenient to study the mass matrices in appendices B and C, respectively.
 
 Inputting the experimental data of two neutrino mass squared differences
 and the mixing angles $\theta_{12}$, $\theta_{23}$ and $\theta_{13}$,
 we can predict the Dirac phase and the Majorana phases. Furthermore, we predict 
the magnitude of the effective neutrino mass of the  $0\nu \beta \beta$,
which is correlated with the Dirac phases.
For the fixed effective neutrino mass for the $0\nu \beta \beta$, there are 
two fold degeneracy and four fold degeneracy
 in  the Majorana phase difference $(\alpha-\beta)$ and  the Dirac phase $\delta_{CP}$, respectively. 
 Therefore, if both $\delta_{CP}$ and $m_{ee}$ are observed, 
the Majorana phases are determined finally.
 At present, the predicted magnitudes of $m_{ee}$ are 
in the regions as $32~\text{meV}\lesssim |m_{ee}|\lesssim 49~\text{meV}$ and 
$34~\text{meV}\lesssim |m_{ee}|\lesssim 59~\text{meV}$ for the cases of $m_{3}=0$ and $m_{3}\not=0$, respectively.
These values are  close to the expected reaches of the coming experiments of the  $0\nu \beta \beta$.

The sum of three neutrino masses is also predicted in both models as
 $0.0952~\text{eV}\lesssim \sum m_i\lesssim 0.101~\text{eV}$ and $0.150~\text{eV}\lesssim \sum m_i\lesssim 0.160~\text{eV}$, respectively.
Those total masses are also within the reaches of the future cosmological and astrophysical measurements.

\vspace{0.5cm}
\noindent
{\bf Acknowledgement}

This work is supported by JSPS Grant-in-Aid for Scientific Research, 15K05045.

\appendix 

\section*{Appendix}

\section{Multiplication rules of the $S_4$ group}
We show the  multiplication rules of $S_4$:
\begin{align}
\begin{pmatrix}
a_1 \\
a_2
\end{pmatrix}_{\bf 2} \otimes 
\begin{pmatrix}
b_1 \\
b_2
\end{pmatrix}_{\bf 2}
&= (a_1b_1+a_2b_2)_{{\bf 1}} \oplus (-a_1b_2+a_2b_1)_{{\bf 1}'} \oplus 
\begin{pmatrix}
a_1b_2+a_2b_1 \\
a_1b_1-a_2b_2
\end{pmatrix}_{{\bf 2}\ ,} \nonumber \\
\begin{pmatrix}
a_1 \\
a_2
\end{pmatrix}_{\bf 2} \otimes 
\begin{pmatrix}
b_1 \\
b_2 \\
b_3
\end{pmatrix}_{{\bf 3}}&= 
\begin{pmatrix}
a_2b_1 \\
-\frac{1}{2}(\sqrt 3a_1b_2+a_2b_2) \\
\frac{1}{2}(\sqrt 3a_1b_3-a_2b_3)
\end{pmatrix}_{{\bf 3}} \oplus 
\begin{pmatrix}
a_1b_1 \\
\frac{1}{2}(\sqrt 3a_2b_2-a_1b_2) \\
-\frac{1}{2}(\sqrt 3a_2b_3+a_1b_3)
\end{pmatrix}_{{\bf 3}'\ ,} \nonumber \\
\begin{pmatrix}
a_1 \\
a_2
\end{pmatrix}_{\bf 2} \otimes 
\begin{pmatrix}
b_1 \\
b_2 \\
b_3
\end{pmatrix}_{{\bf 3}'}&= 
\begin{pmatrix}
a_1b_1 \\
\frac{1}{2}(\sqrt 3a_2b_2-a_1b_2) \\
-\frac{1}{2}(\sqrt 3a_2b_3+a_1b_3)
\end{pmatrix}_{{\bf 3}} \oplus 
\begin{pmatrix}
a_2b_1 \\
-\frac{1}{2}(\sqrt 3a_1b_2+a_2b_2) \\
\frac{1}{2}(\sqrt 3a_1b_3-a_2b_3)
\end{pmatrix}_{{\bf 3}'\ ,} \nonumber \\
\begin{pmatrix}
a_1 \\
a_2 \\
a_3
\end{pmatrix}_{{\bf 3}} \otimes 
\begin{pmatrix}
b_1 \\
b_2 \\
b_3
\end{pmatrix}_{{\bf 3}} &= (a_1b_1+a_2b_2+a_3b_3)_{{\bf 1}} \oplus 
\begin{pmatrix}
\frac{1}{\sqrt 2}(a_2b_2-a_3b_3) \\                                            
\frac{1}{\sqrt 6}(-2a_1b_1+a_2b_2+a_3b_3)
\end{pmatrix}_{\bf 2} \nonumber \\
&\ \oplus 
\begin{pmatrix}
a_2b_3+a_3b_2 \\
a_1b_3+a_3b_1 \\
a_1b_2+a_2b_1
\end{pmatrix}_{{\bf 3}} \oplus 
\begin{pmatrix}
a_3b_2-a_2b_3 \\
a_1b_3-a_3b_1 \\
a_2b_1-a_1b_2
\end{pmatrix}_{{\bf 3}'\ ,} \nonumber \\
\begin{pmatrix}
a_1 \\
a_2 \\
a_3
\end{pmatrix}_{{\bf 3}'} \otimes 
\begin{pmatrix}
b_1 \\
b_2 \\
b_3
\end{pmatrix}_{{\bf 3}'} &= (a_1b_1+a_2b_2+a_3b_3)_{{\bf 1}} \oplus 
\begin{pmatrix}
\frac{1}{\sqrt 2}(a_2b_2-a_3b_3) \\                                            
\frac{1}{\sqrt 6}(-2a_1b_1+a_2b_2+a_3b_3)
\end{pmatrix}_{\bf 2} \nonumber \\
&\ \oplus 
\begin{pmatrix}
a_2b_3+a_3b_2 \\
a_1b_3+a_3b_1 \\
a_1b_2+a_2b_1
\end{pmatrix}_{{\bf 3}} \oplus 
\begin{pmatrix}
a_3b_2-a_2b_3 \\
a_1b_3-a_3b_1 \\
a_2b_1-a_1b_2
\end{pmatrix}_{{\bf 3}'\ ,} \nonumber \\
\begin{pmatrix}
a_1 \\
a_2 \\
a_3
\end{pmatrix}_{{\bf 3}} \otimes 
\begin{pmatrix}
b_1 \\
b_2 \\
b_3
\end{pmatrix}_{{\bf 3}'} &= (a_1b_1+a_2b_2+a_3b_3)_{{\bf 1}'} \oplus 
\begin{pmatrix}
\frac{1}{\sqrt 6}(2a_1b_1-a_2b_2-a_3b_3) \\
\frac{1}{\sqrt 2}(a_2b_2-a_3b_3)
\end{pmatrix}_{\bf 2} \nonumber \\
&\ \oplus 
\begin{pmatrix}
a_3b_2-a_2b_3 \\
a_1b_3-a_3b_1 \\
a_2b_1-a_1b_2
\end{pmatrix}_{{\bf 3}} \oplus 
\begin{pmatrix}
a_2b_3+a_3b_2 \\
a_1b_3+a_3b_1 \\
a_1b_2+a_2b_1
\end{pmatrix}_{{\bf 3}'\ .}
\end{align}
More details are shown in the review~\cite{Ishimori:2010au,Ishimori:2012zz}.

\newpage 

\section{Neutrino mass matrix in the bimaximal base}

\subsection{$m_3=0$ case}
The left-handed Majorana neutrino mass matrix $M_\nu $ is
\begin{equation}
M_\nu =
\begin{pmatrix}
a+b & b & b \\
b & b & b \\
b & b & b
\end{pmatrix}. 
\end{equation} 
After rotating $M_\nu$ with the  bimaximal mixing matrix $V_\text{BM}$ as 
\begin{equation}
V_\text{BM}=
\begin{pmatrix}
\frac{1}{\sqrt{2}} & \frac{1}{\sqrt{2}} & 0 \\
-\frac{1}{2} & \frac{1}{2} & -\frac{1}{\sqrt{2}} \\
-\frac{1}{2} & \frac{1}{2} & \frac{1}{\sqrt{2}}
\end{pmatrix},
\end{equation}
the left-handed Majorana neutrino mass matrix is rewritten as
\begin{equation}
M_\nu =V_\text{BM}
\begin{pmatrix}
\frac{1}{2}\left (a+(3-2\sqrt{2})b\right ) & \frac{1}{2}(a-b) & 0 \\
\frac{1}{2}(a-b) & \frac{1}{2}\left (a+(3+2\sqrt{2})b\right ) & 0 \\
0 & 0 & 0
\end{pmatrix}V_\text{BM}^T~.
\end{equation}

\subsection{$m_3\not =0$ case}
The left-handed Majorana neutrino mass matrix $M_\nu $ is
\begin{equation}
M_\nu =
\begin{pmatrix}
b & b & b \\
b & 2a+b & a+b \\
b & a+b & 2a+b
\end{pmatrix}. 
\end{equation} 
After rotating $M_\nu$ with the  bimaximal mixing matrix $V_\text{BM}$, 
the left-handed Majorana neutrino mass matrix is rewritten as
\begin{equation}
M_\nu =V_\text{BM}
\begin{pmatrix}
\frac{1}{2}\left (3a+(3-2\sqrt{2})b\right ) & -\frac{1}{2}(3a+b) & 0 \\
-\frac{1}{2}(3a+b) & \frac{1}{2}\left (3a+(3+2\sqrt{2})b\right ) & 0 \\
0 & 0 & a
\end{pmatrix}V_\text{BM}^T~.
\end{equation}

\newpage
\section{Neutrino mass matrix in the  $\mu $-$\tau $ maximal base}

\subsection{$m_3=0$ case}
The left-handed Majorana neutrino mass matrix $M_\nu $ is
\begin{equation}
M_\nu =
\begin{pmatrix}
a+b & b & b \\
b & b & b \\
b & b & b
\end{pmatrix}. 
\end{equation} 
After rotating $M_\nu$ with the $\mu $-$\tau $ maximal mixing matrix $V_{23}$ as 
\begin{equation}
V_{23}=
\begin{pmatrix}
1 & 0 & 0 \\
0 & \frac{1}{\sqrt{2}} & -\frac{1}{\sqrt{2}} \\
0 & \frac{1}{\sqrt{2}} & \frac{1}{\sqrt{2}}
\end{pmatrix},
\end{equation}
the left-handed Majorana neutrino mass matrix is rewritten as
\begin{equation}
M_\nu =V_{23}
\begin{pmatrix}
a+b & \sqrt{2}b & 0 \\
\sqrt{2}b & 2b & 0 \\
0 & 0 & 0
\end{pmatrix}V_{23}^T~.
\end{equation}

\subsection{$m_3\not =0$ case}
The left-handed Majorana neutrino mass matrix $M_\nu $ is
\begin{equation}
M_\nu =
\begin{pmatrix}
b & b & b \\
b & 2a+b & a+b \\
b & a+b & 2a+b
\end{pmatrix}. 
\end{equation} 
After rotating $M_\nu$ with the $\mu $-$\tau $ maximal mixing matrix $V_{23}$, 
the left-handed Majorana neutrino mass matrix is rewritten as
\begin{equation}
M_\nu =V_{23}
\begin{pmatrix}
b & \sqrt{2}b & 0 \\
\sqrt{2}b & 3a+2b & 0 \\
0 & 0 & a
\end{pmatrix}V_{23}^T~.
\end{equation}

\end{document}